\documentclass[usenatbib]{mnras}

\usepackage{newtxtext,newtxmath}

\usepackage{ae,aecompl}

\usepackage{graphicx}	
\usepackage{amsmath}	
\usepackage{amssymb}	
\usepackage{multicol}
\usepackage{hyperref}

\title[ELAIS N1 GWB]{Detailed study of ELAIS N1 field with the uGMRT - II.
Source Properties and Spectral Variation Of Foreground Power Spectrum
from 300-500 MHz Observations}

\author[A Chakraborty et al.]{
Arnab Chakraborty,$^{1}$\thanks{E-mail: phd1601121009@iiti.ac.in}
Nirupam Roy, $^{2}$
Abhirup Datta,$^{1}$
Samir Choudhuri $^{4}$
\newauthor
Kanan K. Datta,$^{5}$
Prasun Dutta, $^{3}$
Somnath Bharadwaj, $^{7}$
Huib Intema$^{6}$
\newauthor
Madhurima Choudhury,$^{1}$
Srijita Pal,$^{7}$
Tirthankar Roy Choudhury $^{4}$ 
\\
$^{1}$Discipline of Astronomy, Astrophysics and Space Engineering, Indian Institute of Technology Indore, Indore 453552, India\\
$^{2}$Department of Physics , Indian Institute of Science,Bangalore 560012,India\\
$^{3}$Department of Physics, IIT (BHU) Varanasi, 221005 India\\
$^{4}$National Centre For Radio Astrophysics,Post bag 3,Ganeshkhind,Pune 411007,India\\
$^{5}$Department of Physics, Presidency University, 86/1 College Street, Kolkata-700073, India\\
$^{6}$International Centre for Radio Astronomy Research -- Curtin University, GPO Box U1987, Perth, WA 6845, Australia\\
$^{7}$Department of Physics $\&$ Centre for Theoretical Studies,IIT Kharagpur,Kharagpur 721302,India\\
}
\date{Accepted XXX. Received YYY; in original form ZZZ}

\pubyear{2015}

\begin{document}
\label{firstpage}
\pagerange{\pageref{firstpage}--\pageref{lastpage}}
\maketitle

\begin{abstract}
Understanding the low-frequency radio sky in depth is necessary to subtract foregrounds in order to detect the redshifted 21 cm signal of neutral hydrogen  from the Cosmic Dawn, Epoch of Reionization (EoR) and post-reionization
era.  In this second paper of the series, we present the upgraded Giant Metrewave Radio Telescope (uGMRT)  observation of the ELAIS N1 field made at 300-500 MHz. The image covers an area of $\sim 1.8$ $\mathrm{deg}^{2}$ and has a central background rms noise of  $\sim$ 15 $\mu \mathrm{Jy}$ $\mathrm{beam}^{-1}$. We present a radio source catalog containing 2528 sources (with flux densities > 100 $\mu$Jy) and normalized source counts derived from that. The detailed comparison of detected sources with previous radio observations is shown. We discuss  flux scale accuracy,  positional offsets, spectral index distribution and  correction factors in source counts.  The normalized source counts are in agreement with previous observations of the same field, as well as model source counts from the Square Kilometre Array Design Study (SKADS) simulation. It shows a flattening below $\sim$1 mJy which corresponds to rise in  population of star forming galaxies and radio-quiet AGN. For the first time, we estimated the spectral characteristics of the angular power spectrum or Multi-Frequency Angular Power Spectrum (MFAPS) of diffuse Galactic synchrotron emission (DGSE) over the wide frequency bandwidth  of $300-500$~MHz from radio interferometric observations. This work  demonstrates the improved capabilities of the uGMRT.
\end{abstract}




\section{Introduction}

In the structure formation history of the Universe, after the epoch of recombination at $\textit{z} \sim 1100 $, the Universe was completely neutral and  consisted mostly of neutral hydrogen (HI). In the absence of any radiating sources, the Universe entered to an era known as the  cosmic `Dark Ages'. The formation of first stars and galaxies inaugurated another phase of the Universe, the so-called `Cosmic Dawn' (CD) era spanning the redshift range  $30>\textit{z}>12$. The high-energy photon emanating from the first stars and quasars began to heat and ionize HI in the surrounding Inter-Galactic Medium (IGM) and forced the Universe to go through a patchy phase transition from being fully neutral to completely ionized. This epoch is marked as `Epoch of Reionization' (EoR) spanning the redshift range $12 > \textit{z} > 6$ (For more details see: \citealt{Madau1997,Furlanetto2006,Pritchard2012,Barkana2016,Dayal2018}).

The redshifted 21 cm line signal generated by the hyperfine transition of neutral hydrogen atom (HI) in the IGM is an excellent probe of the early Universe at $\textit{z} > 6$ \citep{Field1958,Madau1997,Furlanetto2006,Pritchard2012}.  Several low-frequency radio interferometers such as the  Donald C.Backer Precision Array to Probe the Epoch of Reionization (PAPER, \citealt{Parsons2014}), the low-frequency Array (LOFAR, \citealt{vanHaarlem2013}), the Murchison Wide -field Array (MWA, \citealt{Li2018}) and the Hydrogen Epoch of Reionization Array (HERA, \citealt{Neben2016,DeBoer2017}), are trying  to measure fluctuations in the cosmological 21 cm signal by means of power spectra. Upcoming instruments such as the Square Kilometre Array (SKA) will have enough sensitivity to resolve physical scales down to 5-10  Mpc (comoving) in the sky plane and corresponding physical scale along line of sight  at $\textit{z} \sim 6-10 $ allowing for tomographic imaging of the 21 cm signal \citep{Koopmans2015,Mondal2018,Mondal2019}.

In addition to these, measuring brightness temperature fluctuations of 21 cm signal in the post EoR era ($\textit{z}<6$) is a promising tool to study the large scale structure of the Universe in three dimensions. This novel technique is widely known as HI intensity mapping.  HI is a biased tracer of dark matter density field. Hence, measurement of the post EoR power spectrum can be used to  study large scale structure formation, Baryon Accoustic Oscillation (BAO), neutrino mass, source clustering, etc \citep{Bharadwaj2001a,Bharadwaj2003,Bharadwaj2005,Wyithe2008,Visbal2009,Bull2015,Santos2015}. There are some instruments such as  BAOBAB \citep{Pober2013a}, BINGO \citep{Battye2012}, CHIME \citep{Bandura2014}, the Tianlai project \citep{Chen2016}, HIRAX \citep{Newburgh2016}, SKA1-MID \citep{Bull2015}, OWFA \citep{Subrahmanya2017} trying to measure   Baryon Acooustic Oscillations (BAO) over a redshift range  z $\sim 0.5 - 2.5 $, which can be used as a standard ruler to constrain the Dark Energy equation of state. 
  
The main challenge to detect the cosmological HI signal, common to all of these experiments, is the strong contamination of systematic effects (ionospheric distortion, telescope response, calibration, etc) and bright foregrounds (Galactic and extragalactic) \citep{Datta2009}. Foreground sources include  diffuse Galactic synchrotron emission from our Galaxy (DGSE) \citep{Shaver1999}, free-free emission from Galactic and extragalactic sources \citep{Cooray2004}, faint radio-loud quasars \citep{Di Matteo2002}, synchrotron emission from low-redshift Galaxy clusters \citep{Di Matteo2004}, extragalactic point sources,  etc.  Typically, foregrounds are 4-5 orders of magnitude stronger than the  redshifted HI signal \citep{Zaldarriag2004,Bharadwaj2005,Jeli&cacute;2008,Bernardi2009,Jeli&cacute;2010,Zaroubi2012,Chapman2015}. There are several different ways to deal with  foregrounds, but all the methods rely on the fact that foreground sources to have a smooth spectral shape. However, the redshifted HI 21 cm signal has  spectral structure \citep{Pritchard2012}. In fact, this difference in spectral properties between the strong foreground and faint 21 cm signal can be used favourably for the detection of the cosmological signal \citep{Datta2010}. Hence, the accuracy in the knowledge of the spectral ``smoothness'' of the foreground becomes critical. Our current study makes an attempt to constrain the spectral behaviour of the foregrounds near the redshifted 21 cm signal frequencies. 
The three main techniques proposed to overcome foreground contamination are  foreground avoidance, foreground removal and foreground suppression. 
Instead of isotropic 1D power-spectrum, P($\textit{k}$), of HI brightness temperature fluctuation, cylindrical (2D) power spectrum, P($\textit{k}_{\perp},\textit{k}_{\parallel}$) is a useful diagnostic in terms of foreground avoidance.  Spectral smoothness of foregrounds confines the majority of foreground power to low $\textit{k}_{\parallel}$  modes, resulting  in ``Foreground Wedge''.  In foreground avoidance technique, the EoR signal is searched for outside this wedge, in so called ``EoR Window'' \citep{Datta2010,Parsons2012,Vedantham2012,Pober2013,Thyagarajan2013,Dillon2015}. However, error in calibration of chromatic instrument and insufficient knowledge on wedge boundary can leak foreground power into the wedge and consequently detection of 21 cm signal becomes challenging even inside the EoR window. 
Foregrounds can be modelled very precisely and subtracted from the data set. Also, without any modelling a component analysis method can be used to mitigate foregrounds (For detail see: \citealt{Chapman2012,Chapman2013}).  The foregrounds can be suppressed by weighting foregrounds dominated modes appropriately \citep{Liu2011}.

The power spectrum of DGSE is generally modelled as a power law both as a function of frequency and angular scale  \citep{Santos2005,Datta2007}  : 
\begin{equation}
\mathcal{C}_{\mathcal{\ell}}(\nu) = A \Big(\frac{\mathcal{\ell}}{\mathcal{\ell}_{0}}\Big) ^{ -\beta} \Big(\frac{\nu}{\nu_{0}}\Big)^{-2\alpha}, 
\label{power_law_model}
\end{equation}
where $\beta$ is the power law index of the angular power spectrum (APS) of DGSE and $\alpha$ is the mean  spectral index.
There are several observational constraint on angular fluctuation in DGSE   \citep{Ali2008,Iacobelli2013, Bernardi2009, Ghosh2012, Iacobelli2013,Choudhuri2017,Chakraborty2019}. Motivated by power law nature of synchrotron emission, spectral evolution of fluctuation in DGSE is also modelled as a power law.  Basic principle of  foreground subtraction technique is to fit this simple power law model for DGSE  along frequency axis for each pixel of a data cube and subtract it from the data. However, constraint on  spectral variation of foreground power spectrum based on  observation is necessary to model DGSE in a most precise manner. As, any error in foreground subtraction  can remove the  whole 21 cm signal from the data set, accurate modelling is crucial (see Sec. \ref{sec.DGSE}).   

 In this aspect, deep understanding of low-frequency radio sky is important to model extragalactic and Galactic foregrounds. We need prior knowledge in  Spectral Energy Distribution (SED), clustering and evolutionary properties of extragalactic sources in radio band to achieve accurate modelling of foregrounds \citep{Jeli&cacute;2008,Jeli&cacute;2010,Prandoni2018}. Spatial distribution of sources, in general,  is assumed to be Poissonian  or a very simple clustering with a power law feature \citep{Ali2008,Jeli&cacute;2010,Trott2016}. Source counts are also assumed to follow a single power law \citep{Intema2017,Hurley-Walker2017,Franzen2019}. But, several studies show a deviation from this single power law model  at sub-mJy to $\mu \mathrm{Jy}$ flux densities \citep{Williams2016,Prandoni2018,Hale2019}.  Any error in modelling of foregrounds can be harmful in 21 cm signal detection.  

In addition, differential source counts can give constraint on the nature  of extragalactic sources.  We have good understanding of source population at high flux densities for 1.4 GHz observations. Source counts below 1 mJy is a subject of much debate.  Several low-frequency deep survey found a flattening in normalized source counts around 1 mJy \citep{Windhorst1985,Williams2016,Prandoni2018A,Hale2019}. This suggests a increase in population of Star Forming Galaxies (SFG), radio-quiet AGN  at low flux densities \citep{Jackson1999,Prandoni2018A}. Also, detection of SFGs and AGNs through their radio emission at low-frequency together with their redshift information will help us to understand astrophysical properties of these sources such as luminosity, size of the source, cosmic-ray electron population, etc \citep{Simpson2017,Norris2017}. 
Our knowledge of low-frequency sky is poor compared with that of $\gtrsim ~1.4$~GHz and as a consequence empirical constraint on low-frequency source count is limited. This is  mainly because of the fact that reaching high SNR  at low-frequency is extremely challenging.

So, deep survey at low-frequency is important not only for modelling foregrounds to detect  cosmological 21 cm signal, but it is equally  useful to understand  astrophysical properties  of extragalactic sources.   In our first paper of this series \citep{Chakraborty2019}, we have shown the angular power spectrum of Galactic and extragalactic foregrounds. The spatial behavior of DGSE in this field has been quantified. We have also shown, the effect of direction-dependent and direction-independent calibration in estimation of angular power spectrum of DGSE  in this field using   32 MHz (GSB: GMRT Software Backend) bandwidth data of the ELAIS N1 field. In this second paper of the  series, we  present the detailed study of the ELAIS N1 field  with wide bandwidth data (300-500 MHz) using the uGMRT.   ELAIS N1 field has been previously studied at other frequencies \citep{Ciliegi1999,Garn2008,Sirothia2009,Jelic2014,Taylor2016}. The NVSS \citep{Condon1998} and FIRST \citep{Becker1995} surveys both cover the ELAIS N1 region, but only to relatively shallow 5$\sigma$ limits of 2.25 and 0.75 mJy respectively.   

Although, this field has been studied in different frequencies, the information at low-frequency with high resolution is still lacking. We have observed the ELAIS N1 field at Band-3 (300 - 500 MHz) using uGMRT in GTAC (GMRT Time Allocation Committee) cycle-32 for 25 hours to get a deep high resolution map of the same field.  We present $\sim$ 1.8 $\mathrm{deg}^{2}$ map of  ELAIS N1 with an rms depth of  $\sim$ 15 $\mu \mathrm{Jy}$ per beam.  We have created a source catalog, down to 100 $\mu$Jy flux density, consisting of 2528 sources. A detailed comparison of our  catalog with other radio frequency observations is also shown. We have  estimated normalized Euclidian source counts and compared this with previous  observation of the same field. To model fluctuation in DGSE as a function of frequency, we have estimated  the spectral variation of  angular power spectrum of DGSE  with Tapered Gridded Estimator (TGE) using the whole bandwidth (200 MHz) data.    In our fourthcoming papers we will present  the clustering properties of  sources in this field,  3D power spectrum to get upper limit on the post-EoR signal, effect of calibration on the ``Foreground Wedge'', etc.

The paper structured as follows. In Sec. \ref{sec.observation}, we describe uGMRT  Band-3 (300-500 MHz) observation of  ELAIS N1 field.  Details of data reduction including flagging, calibration, imaging and self-calibration are mentioned in Sec. \ref{sec.data_reduction}. In Sec. \ref{sec.source_catalog}, we present source catalog; detailed comparison with other catalogs at different frequencies. The normalized source counts along with the correction factors are described in Sec. \ref{sec.source_counts}. In Sec. \ref{sec.DGSE}, we have shown the evolution of spectral index of DGSE power spectrum across the whole band. Finally, we draw conclusion in Sec. \ref{conclusion}.

\section{Observation}
\label{sec.observation}

\begin{figure*}
     \centering
     \begin{tabular}{c|c}
     \includegraphics[width=3.2in,height=2.8in]{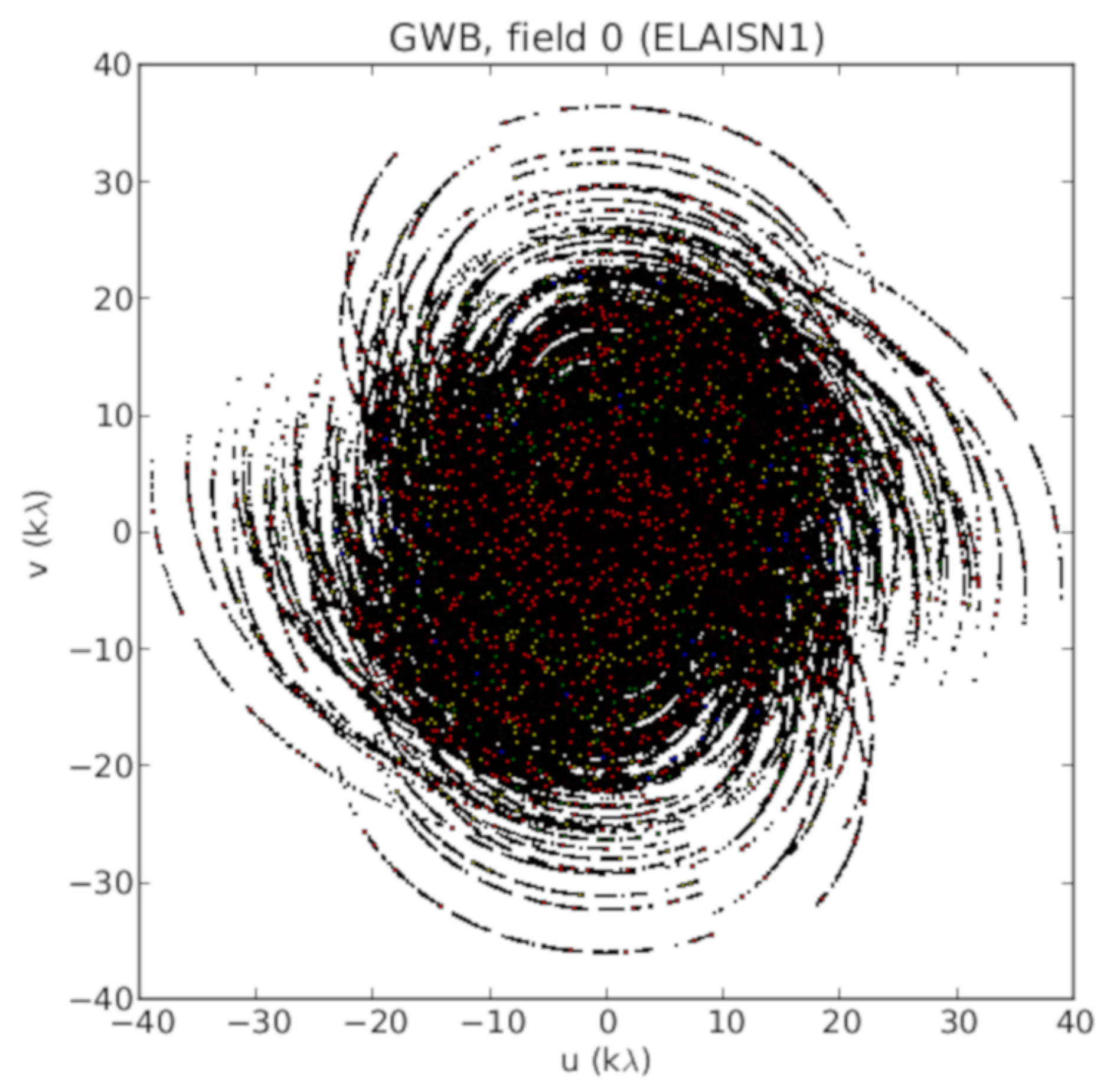} &
     \includegraphics[width=3.2in,height=2.8in]{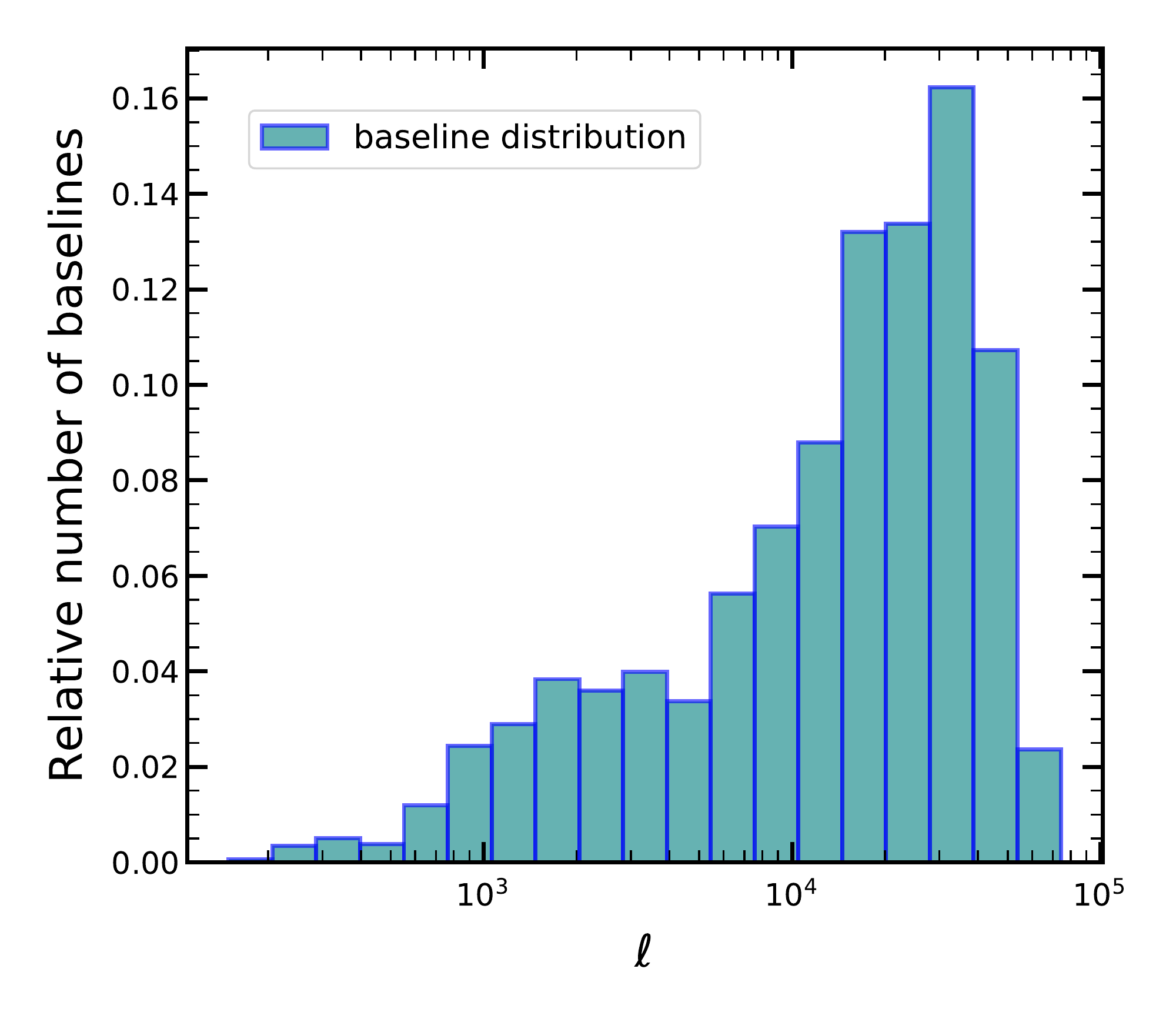}
     \end{tabular}
     \caption{Left panel: The \textit{uv}-coverage of ELAIS N1 field in k$\lambda$ using the uGMRT for 300-500 MHz bandwidth. Only 6\% of the total data points has been plotted. Large bandwidth and long observational time results in  a densely filled \textit{uv}-plane.  Right panel: The relative baseline distribution as a  function $\mathcal{\ell}$, where $\mathcal{\ell}$ = $2\pi$U. U is the baseline length. This illustrates the sensitivity of uGMRT at different angular scale to estimate the APS (see Sec. \ref{sec.DGSE}).}
     \label{uv}
 \end{figure*}

We carried out deep observation of the ELAIS N1 field ($\alpha_{2000}=16^{h}10^{m}1^{s} ,\delta_{2000}=54^{\circ}30'36\arcsec$ ) with uGMRT in GTAC cycle 32 during May 2017. The total observation time was  25 hours (including calibrators) spanned  over four days. The ELAIS N1 field is at high Galactic latitudes ($\textit{l}=86.95^{\circ} ,b=+44.48^{\circ}$). During the GTAC cycle 32, ELAIS N1 field was up at night time and the  observation was carried out at night for all four  days. 
The observation spanned a frequency range of 300-500 MHz, i.e, bandwidth (BW) is 200 MHz. The whole band was divided into 8192 channels, resulting in frequency resolution of 24 KHz. The integration time used was 2s.
This high time and frequency resolution was helpful in identifying and flagging RFI (see Sec. \ref{flagging}). The \textit{uv}-coverage of ELAIS N1 field, using uGMRT, can be seen in the left panel of Fig. \ref{uv}. We have a  densely filled  core of the  \textit{uv}-plane. The large bandwidth fills the \textit{uv}-plane radially and long observational time fills the plane azimuthally using Earth's rotational speed.  However, gaps  in the  \textit{uv}-plane  limits the quality of the final image.  In the right panel of Fig. \ref{uv}, we show the relative number of baseline distribution at different multipole ($\mathcal{\ell}$ = $\mathrm{U}/2\pi$) to demonstrate the sensitivity of the uGMRT at different angular scale in the estimation of the angular power spectrum (see Sec. \ref{sec.DGSE}).

 We have observed  a flux calibrator 3C286 in the beginning and  in the middle of each observing session. We have also observed 3C48 at the end since 3C286 was not up during the last scan of the observation.  We have observed a Phase calibrator J1549+506 near  the target field    every 25 minutes in between scans on the target field.  The total on-source data after exclusion of calibrators scans is $\sim$ 13 hours.  The observation summary along with the calibrators used is presented in Table \ref{observation}.

\begin{table}
\caption{Observational details of the target field ELAIS N1 and the calibrator sources for four observing sessions}
	
\begin{tabular}[width=\columnwidth]{ll}
\hline
\hline
Project code & 32\_120 \\
Observation date & 5, 6, 7 May 2017 \\
                 &  27 June 2017\\
\hline
Bandwidth &  200 MHz\\
Frequency range & 300-500 MHz\\
Channels & 8192\\
Integration time & 2s\\
Correlations & RR RL LR LL\\
Total on-source time & 13 h (ELAIS N1)\\
Working antennas & 26 \\
\hline
Pointing centres & $13^{h}31^{m}08^{s}$  $+30^{d}30^{m}32^{s}$ (3C286)\\
                 & $15^{h}49^{m}17^{s}$  $+50^{d}38^{m}05^{s}$ (J1549+506)\\
                 & $16^{h}10^{m}01^{s}$  $+54^{d}30^{m}36^{s}$ (ELAIS N1)\\
                 & $01^{h}37^{m}41^{s}$  $+33^{d}09^{m}35^{s}$ (3C48)\\
\hline                 
\hline                 
\end{tabular}
\label{observation}	    
\end{table}

\begin{figure*}
\centering
\includegraphics[width=7.0in]{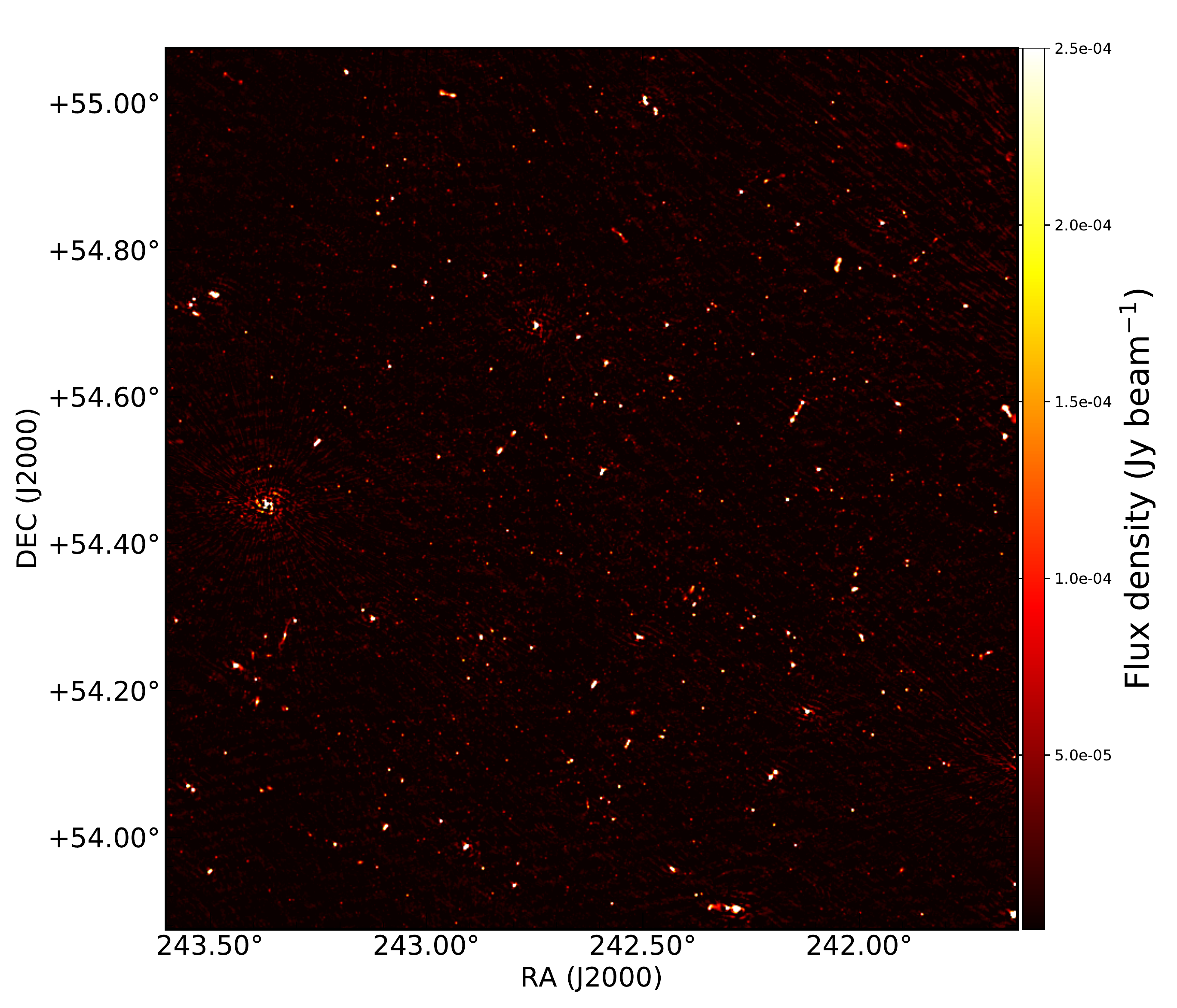}
\caption{The above uGMRT image is zoomed-in total intensity image of ELIAS N1 at 400MHz (bandwidth 200MHz).  The Central off-source noise is $\sim$ 15 $\mu \mathrm{Jy}$ $\mathrm{beam}^{-1}$. The image covers a central area of $\sim$ 1.2 $\mathrm{deg}^{2}$. This illustrates that  a large number of weak sources are detected due to high signal-to-noise ratio achieved here.}
\label{image}
\end{figure*}
\section{Data reduction}
\label{sec.data_reduction}
In this section we describe the RFI mitigation, calibration and imaging  procedures in detail which result in  high fidelity image of the ELAIS N1. 

\subsection{Flagging}
\label{flagging}
Low-frequency radio observation with uGMRT is affected gravely by  man-made radio transmitters. In general, this spurious signal, which is several orders of magnitude stronger than the weak astronomical signal of interest, is known as radio-frequency interference (RFI).   The effect of RFI is particularly strong at frequencies below 600 MHz in uGMRT observation. RFI can be of many forms but in most cases it is localized in frequency domain or it persists for a short time interval (For more detail see \citealt{Offringa2010A}). 
 
 There are numerous techniques available to mitigate RFI from the post-correlation radio data. The widely used method is to identify RFI in high time and frequency resolution and flagged them from the total data set.  Also,  this unfavourable signal may completely corrupt some   baselines or any particular antenna. It is necessary to remove those baselines and any  bad antenna before calibration and imaging. We have taken the data with high time and frequency resolution (24 KHz and 2s) to flag RFI in most efficient way without loosing much of the signal. This results in high data volume (4 TB)  for further processing. We have applied AOFLAGGER \citep{Offringa2012} to the high resolution data set.   It detects anomaly in time-frequency domain per baseline per polarization and flagged them. For more details regarding working methodology of AOFLAGGER please see: \citet{Offringa2010A},\citet{Offringa2010B},\citet{Offringa2012}.
 
\subsection{Calibration}
\label{sec.calibration}
After flagging in high time-frequency resolution we have averaged the data down to 2048 channels and 8s integration time. We have calibrated four night's observations separately but in a consistent manner. The calibration is done with exactly same parameters for different night's data sets.  Then during $\textit{CLEAN}$ing  we use all the calibrated visibilities to make a continuum image. We have not done any polarization calibration as of now. This defers to later work. The data reduction is done using a  {\tiny CASA} \footnote{See: \url{https://casa.nrao.edu}; \citep{McMullin2007} } based pipeline. Here we briefly describe the steps of our data reduction procedures. We have first run Hanning-smoothing to reduce the Gibbs ringing  across frequency channels. Then we have flagged 5\% of total number of channels (2.5\% on each side) from the edge of the bandpass using flagging mode: $\textit{quack}$ in {\tiny CASA}. We have used \citet{Perley2013} to set the  model of our standard primary calibrators 3C286 and 3C48. After setting the flux density model of primary calibrator we follow the traditional direction-independent calibration technique. First we have done a initial delay, gain and bandpass calibration to look for remaining bad data. For gain calibration we have used short solution interval of 16s.   Then we have applied this initial calibration to the primary calibrators and  run $\textit{RFLAG}$ (an automated flagging routine in {\tiny CASA}) on the calibrated data to remove RFI.  We have done this initial calibration followed by flagging with  $\textit{RFLAG}$ for two more times. After this loop, we have done the final delay and bandpass calibration for the primary calibrator. Then we have solved for variation of gains (both amplitude and phase) on a timescale of 16s for primary calibrators and the secondary calibrator (J1549+506) to correct temporal variation. We have used bootstrap method to set the flux density value of the secondary calibrator (J1549+506) using model of the primary calibrator (3C286). Since J1549+506 is close to the target we have transferred the gain solutions from this calibrator to the target (ELAIS N1). Then we have splitted  the target field from the whole data set and proceed to imaging.  The total data flagged,  after flagging (Sec. \ref{flagging}) and calibration of 4 nights data (Sec. \ref{sec.calibration}), is $\sim$ 40\%. 

\subsection{Imaging}
\label{sec.imaging}
After calibration we have used the {\tiny CASA} task $\textit{CLEAN}$ to make a combined continuum image with 4 days data sets. The FoV at 400 MHz of uGMRT is large ($1.15^{\circ} \times 1.15^{\circ}$). Hence, we have taken 256 $\textit{w}$-projection planes to correct for non-coplanar nature of the array. To account for the large bandwidth and spectral structure of the sources present in the field, we have used MS-MFS algorithm \citep{Rau2011} in {\tiny CASA} and choose nterms = 2. We choose Briggs robust parameter = -1, which gives nearly uniform weighting. This particular choice of robust parameter produces a near-Gaussian central PSF and  suppresses the broad wings. We made a large image of size $3^{\circ} \times 3^{\circ}$ to include bright sources lying outside the FoV. Modelling these sources during $\textit{CLEAN}$ing is important, otherwise large sidelobes of those will distort the image within the primary beam area.  
\subsection{Self-Calibration}
After getting the first image, we have performed 4 rounds of phase only self-calibration.  The solution interval of self-calibration loops  are 4min, 2min, 1min and 30sec.   After 4 rounds of self-calibration and imaging loop, the signal-to-noise ratio (SNR) saturates and we got the final image. The off-source rms noise achieved near the centre of the FoV is $\sim$ 15 $\mu$Jy $\mathrm{beam}^{-1}$. The size of the synthesized beam is $4.6\arcsec \times 4.3\arcsec$ and the position angle (pa) of the beam is $-34.2^{\circ}$. The central zoomed-in part of the image is shown in Fig. \ref{image}.

\begin{figure*}
\includegraphics[width=7.0in,height=6.0in]{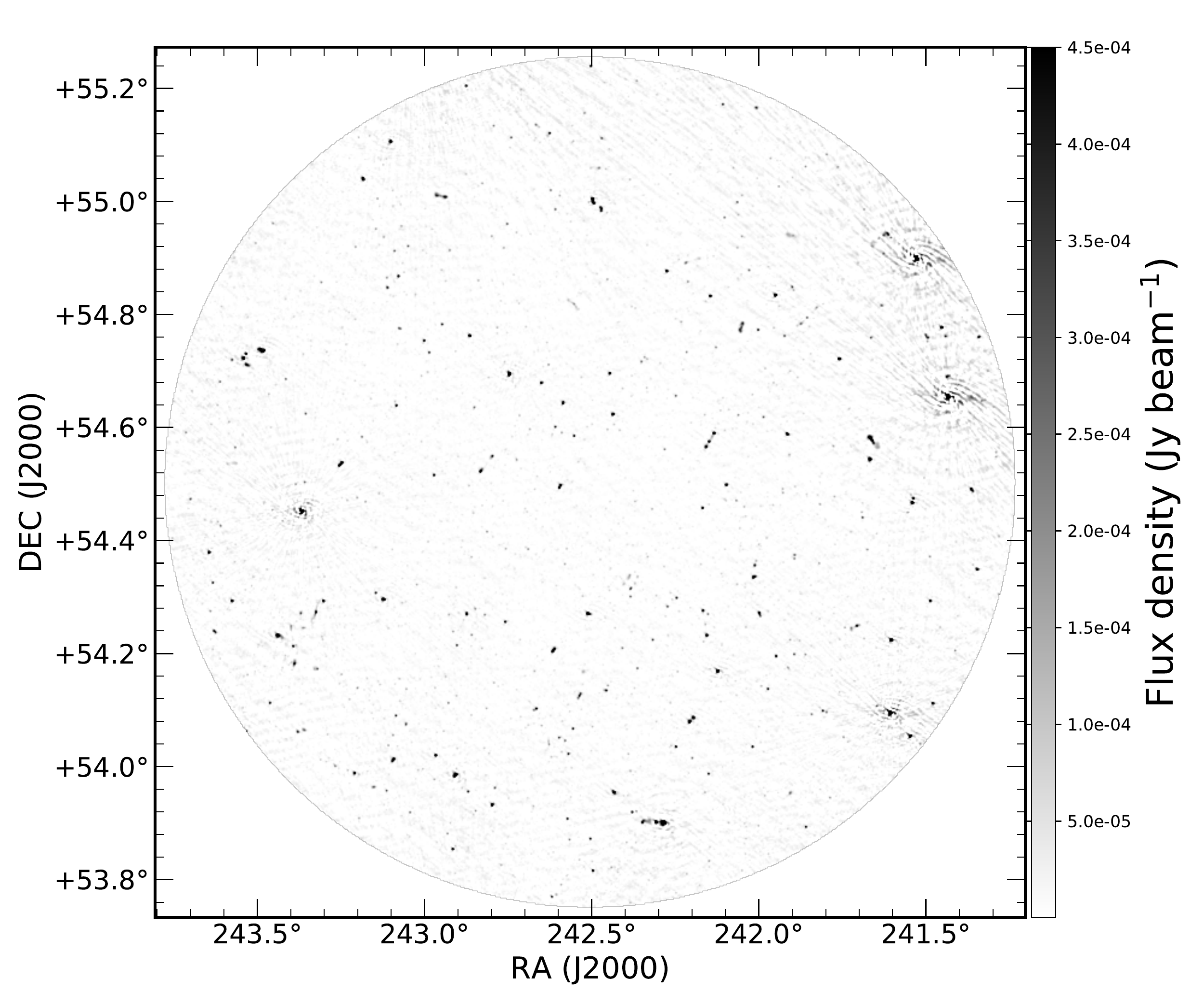}
\caption{Primary beam corrected image of  ELIAS N1 at 400MHz. The image extends over an area of $\sim$ 1.8 $\mathrm{deg}^{2}$. The off source rms at the center is $\sim$ 15 $\mu$Jy $\mathrm{beam}^{-1}$ and beam size is $ 4.6\arcsec \times 4.3\arcsec$.}
\label{PB}
\end{figure*}

We have corrected for the beam model of uGMRT to measure real sky fluxes.  The primary beam of uGMRT is usually modelled as eighth order polynomial. This fitted polynomial is  given by : 
 \begin{equation}
     1+ \Big(\frac{a}{10^{3}}\Big)x^{2} + \Big(\frac{b}{10^{7}}\Big)x^{4} + \Big(\frac{c}{10^{10}}\Big)x^{6} + \Big(\frac{d}{10^{13}}\Big)x^{8} 
 \label{PB_equation}
 \end{equation}
 where, x is in terms of separation from pointing position in arc-minutes times the frequency in GHz and a,b,c,d are the coefficients. For Band-3 (250-500 MHz), the values are:  a = -2.939, b = 33.312, c= -16.659, d = 3.066. We use these primary beam parameters provided by uGMRT staff to make a primary beam corrected map of the ELAIS N1 field. We have imposed a  cut   at 20\% of the peak of the primary beam response. Fig. \ref{PB}  shows the $1.5^{\circ} \times 1.5^{\circ}$ image after primary beam correction.
 
\begin{figure*}
\centering
\begin{tabular}{cc}
\includegraphics[width=\columnwidth]{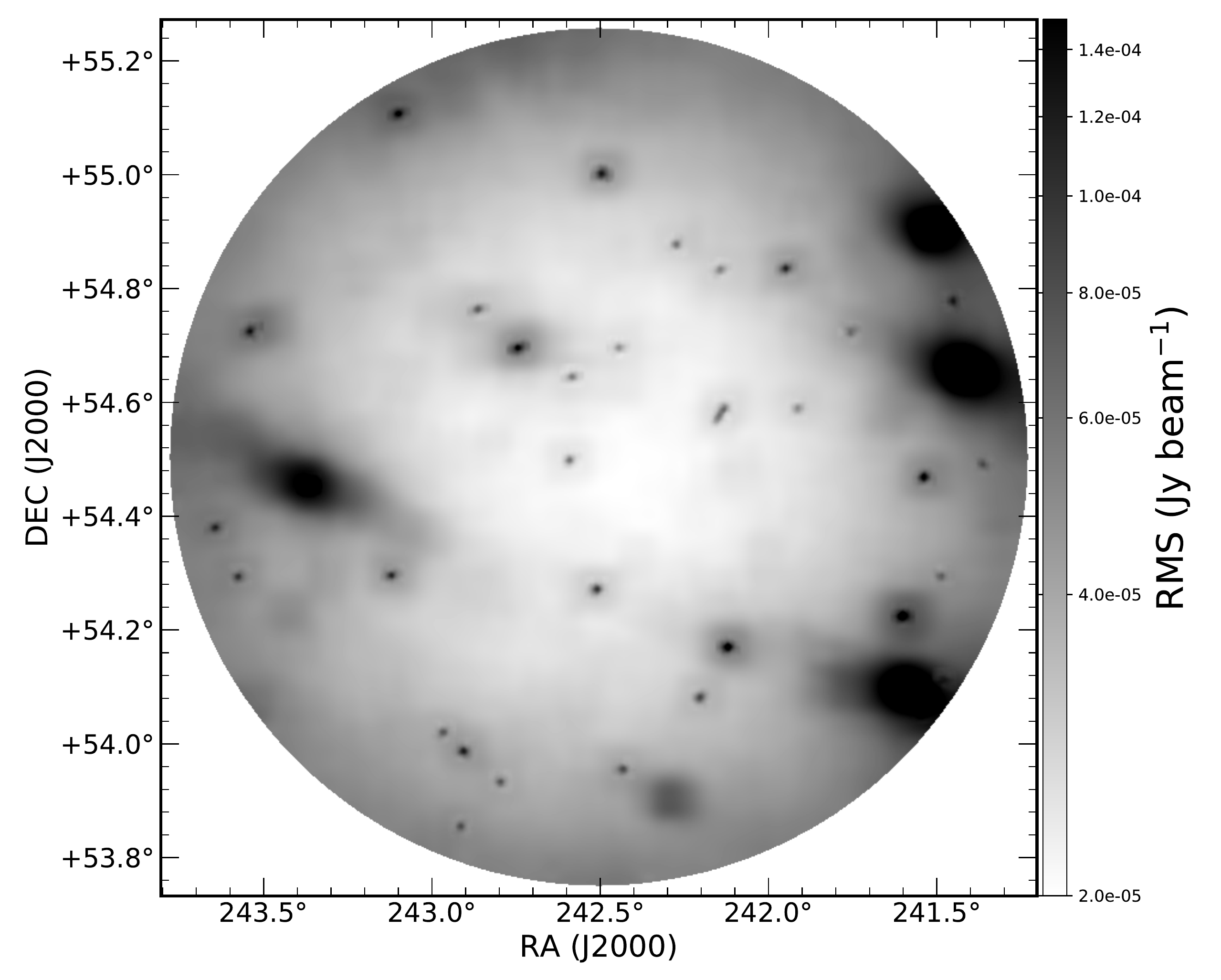}   & \includegraphics[width=\columnwidth]{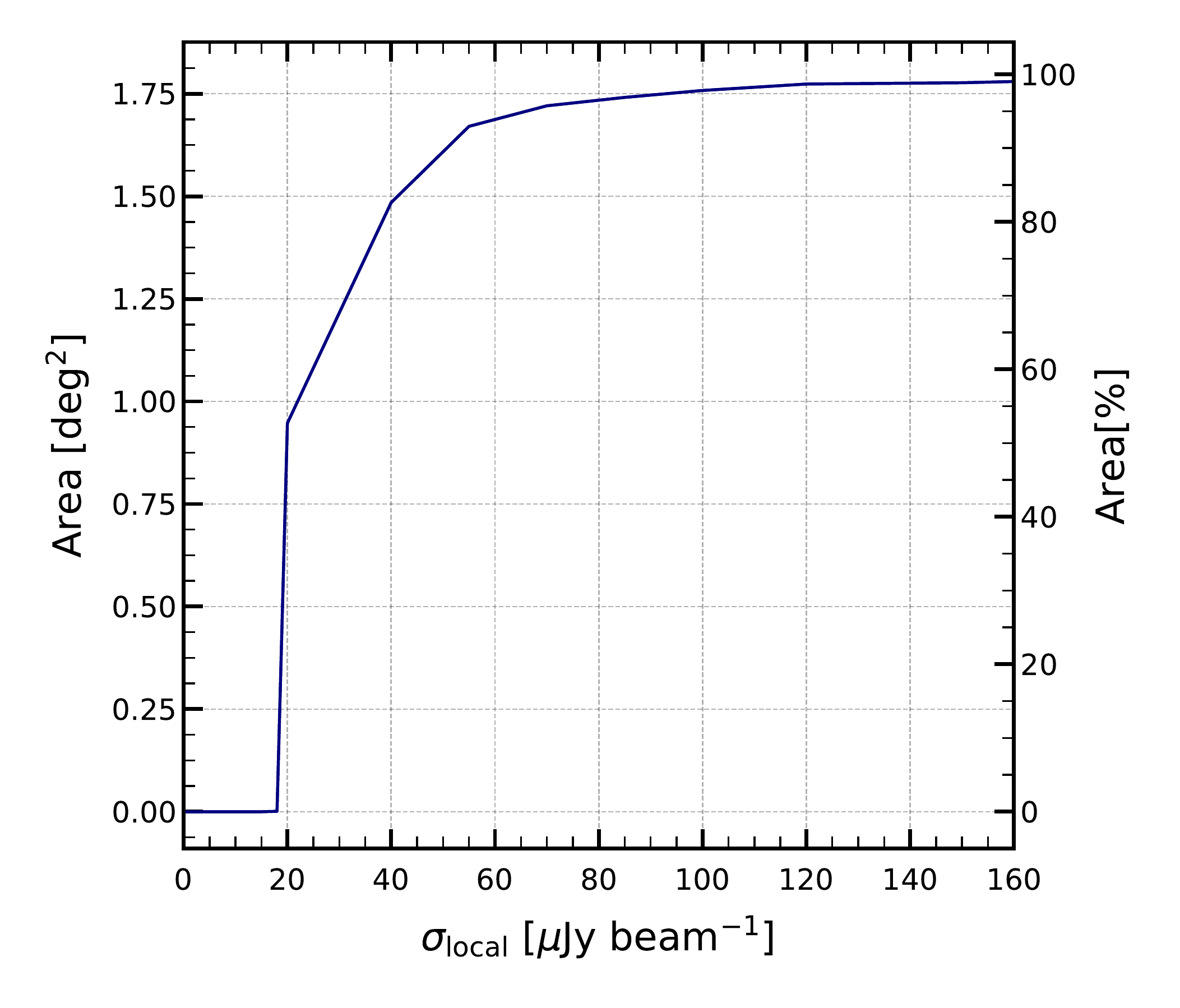}  \\
  
\end{tabular}

\caption{Left image is showing the local rms noise measured in the final map.  Local noise is high near the bright sources and at the edge of FoV. Right: Cumulative area of the final map with a rms noise level below the given value.}
\label{rms}
\end{figure*}

\section{Source Catalog}
\label{sec.source_catalog}
We have assembled  a source catalog using P{\tiny Y}BDSF \citep{Mohan2015} to characterize sources present in this field. Along with source catalog P{\tiny Y}BDSF produces the rms and residual map. The rms map shows the variation of noise across the whole field whereas residual map gives the image of the field with all the modelled sources are subtracted. The left panel of Fig. \ref{rms} shows rms map of the field. It is clear from the image that rms is varying across the FoV with an increased value near the bright sources and at the edge of the FoV. In the right panel of the Fig. \ref{rms}, the area and the corresponding percentage of the image that has a noise value less than a given value is shown.  We have used the primary beam-corrected image to extract the sources with corrected flux values.  P{\tiny Y}BDSF calculates varying rms map across the FoV using a sliding box window rms$\_$box = (180,50) (i.e a box size of 180 pixels in every 50 pixels).  Signal to noise ratio is generally high near bright artifacts. First, we have identified those bright regions whose peak amplitude  are higher than adaptive threshold of 150 $\sigma$, where $\sigma$ is the clipped rms across the entire map. Then to avoid counting artefacts as real sources, we have used a small rms$\_$box = (35,7)  around those bright regions. P{\tiny Y}BDSF identifies islands of contiguous emission over a pixel threshold and then fit multiple Gaussian to each island . We have imposed a threshold of $3\sigma_{\mathrm{rms}}$ to detect islands and pixel threshold of $ 6\sigma_{\mathrm{rms}}$ for source detection.

 The PSF may vary across the FoV due to ionospheric fluctuation  in low-frequency observation. So, at any position in the image the actual PSF is different from the restoring beam by a certain amount. 
 To address this issue, we have calculated variation of PSF using P{\tiny Y}BDSF with $psf\_vary\_do = \mathrm{True}$\footnote{For more details on different parameters please see: \url{https://www.astron.nl/citt/pybdsf/}}. It selects a list of sources which are likely to be unresolved (``S'' flagged sources from PYBDM output) and with high SNR (>$10\sigma$). The number of unresolved sources used are 468. Then using Voronoi tessellation the whole map is tesselated into tiles  around those bright sources and with in which PSF shape has been calculated. The spatial variation of  PSF  is then interpolated across the whole image and the effects of PSF variation are corrected for.

P{\tiny Y}BDSF groups nearby Gaussians within an island into sources. The total flux is obtained by summing the fluxes of grouped Gaussians. The uncertainty in total flux is calculated by summing the Gaussian uncertainties in quadrature.  The source position is set to be its centroid and source size is measured using moment analysis with the knowledge of the image restoring beam. We have also checked the residual rms map and Gaussian model image after fitting to exclude any false detection or a artefacts.  A total of 41 sources has been identified as spurious sources by visual inspection.  They are mainly side-lobes of few bright sources (artefacts),  at the edge of the FoV, detected as real sources by P{\tiny Y}BDSF.  They are residing within 2$\arcmin$ of those bright sources and also do not have any counterpart in high frequency catalog. We have removed those sources from the final catalog.

The number of beams per source is used to determine whether the image is confused or not. If the average number of pixels between two sources are less than 25  then the image is assumed to be confused. Here we got 540 pixels corresponding to nearly 10 beams per source. This ensures that it is not confusion limited.  The confusion noise limit for this observation is $\sim$ 2.09 $\mu$Jy $\mathrm{beam}^{-1}$ \citep{Condon2012}.

\begin{table*}
\caption{Sample of uGMRT 400 MHz source catalog of ELAIS N1 field.}
\label{sample_catalog}
\scalebox{1.0}{
\begin{tabular}{l l l l l l l l l l l l}
\hline
\hline
Id & RA  & E\_RA & DEC & E\_DEC & Total\_flux &  Peak\_flux & Major & Minor & PA & rms  \\

() & (deg) & (arcsec) & (deg) & (arcsec) & (mJy)& (mJy $\mathrm{beam}^{-1}$) & (arcsec) & (arcsec) & (degree) & (mJy $\mathrm{beam}^{-1}$) \\
(1) & (2)  & (3)    & (4) & (5)     & (6) & (7) & (8) & (9) & (10) & (11)\\
 \hline
0  &  243.79 & 0.33 & 54.58 & 0.27 & 1.29 & 0.68 &  6.8 & 5.7 & 65.20 & 0.06 \\
1  &  243.77 & 0.95 & 54.40 & 0.58 & 2.37 & 0.33 &  15.0 & 9.2 & 74.08 & 0.05 \\
3  &    243.78 & 0.31 & 54.64  &  0.38 & 0.79 & 0.47  &  6.7  & 4.8 & 35.81 & 0.06 \\ 
\hline 
\end{tabular}}
\\\flushleft{Notes: The columns of the final catalog (fits format) include source id, positions, error in positions, flux densities, peak flux densities, sizes, position angle and local rms noise.}
\end{table*}

 We have compiled 2528 sources within 20\% of uGMRT primary beam at 400 MHz with flux densities greater than 100 $\mu$Jy ($> 6\sigma$). A sample of the catalog shown in Table \ref{sample_catalog}. The selection of extended and unresolved sources is  discussed below.

\subsection{Classification of sources}

\begin{figure}
\centering
\includegraphics[width=\columnwidth]{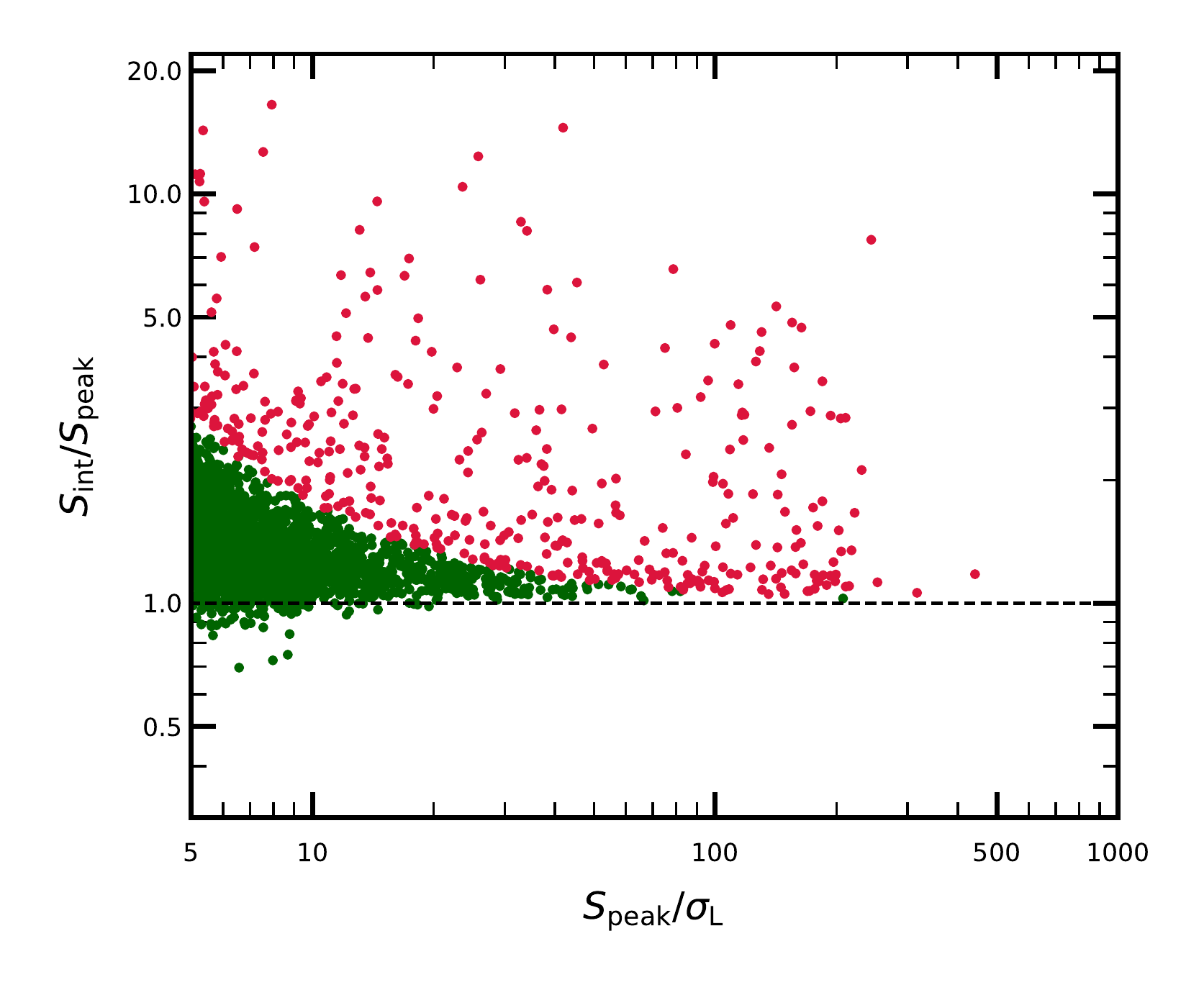}
\caption{The ratio of integrated to peak flux density $\Big(S_{\mathrm{int}}/S_{\mathrm{peak}}\Big)$ as a function of signal-to-noise ratio  $\Big(S_{\mathrm{peak}}$/$\sigma_{\mathrm{L}}\Big)$ of sources. Extended sources are shown in red and point-like sources in green.}
\label{resolved}
\end{figure}

Classification of sources as resolved and point-like is complicated based on P{\tiny Y}BDSF derived source properties. This is mainly due to  time and bandwidth smearing, which artificially extend the sources in the image plane.  The error in calibration  and varying noise  are also responsible to scatter the ratio of integrated flux density ($S_{\mathrm{int}}$) to peak flux density ($S_{\mathrm{peak}}$).  As a consequence of that, we can not simply classify  the sources as extended or resolved based on the requirement of $\Big(S_{\mathrm{int}}/S_{\mathrm{peak}}\Big) > 1$.  In fig. \ref{resolved}, we have plotted $\Big(S_{\mathrm{int}}/S_{\mathrm{peak}}\Big)$ as a function of $\Big(S_{\mathrm{peak}}/\sigma\Big)$, where $\sigma$ is the local rms. The distribution is skewed at low SNR.

The consequence  of bandwidth and time average smearing is the reduction of peak flux densities of the sources whereas the integrated  flux density remains same. As a result the ratio of total to peak flux density is not equals to one for originally unresolved sources. The magnitude of this effect depends on the radial distance from the pointing centre, channel width (frequency resolution) and integration time (time resolution).  We have theoretically estimated the combined effect of bandwidth and time smearing and found that measured peak flux density is 97\% of the expected value (see \citealt{Condon1998}) at the maximum distance (45$\arcmin$) from the phase centre. Hence, smearing is not an issue in our case.

   We have used the method described in \citet{Franzen2015,Franzen2019} for identification of resolved sources.  The requirement for a source to be extended at the 3$\sigma$ level is, following \citet{Franzen2015} : 
\begin{equation}
    \mathrm{ln}\Big(S_{\mathrm{int}}/S_{\mathrm{peak}}\Big) > 3\sqrt{\Big(\frac{\sigma_{\mathrm{S}}}{S_{\mathrm{int}}}\Big)^2 + \Big(\frac{\sigma_{S_{\mathrm{peak}}}}{S_{\mathrm{peak}}}\Big)^2}
\end{equation}
where $\sigma_{\mathrm{S}}$ and $\sigma_{\mathrm{S_{peak}}}$ are the uncertainties on integrated flux density ($S_{\mathrm{int}}$) and peak flux density ($S_{\mathrm{peak}}$) respectively.   We have found 401 (the red circles) resolved sources based on the above requirements and 2127 sources have been classified as unresolved or point-like sources at this frequency of observation with uGMRT (see  Fig. \ref{resolved} ).  

\section{Comparison with other radio catalogs}
In this section, we present detailed comparison with other radio catalogs having overlapping regions. We have compared our catalog with NVSS and FIRST all-sky survey and GMRT observation of the ELAIS N1 field at 325 MHz by \citet{Sirothia2009} and at 610 MHz by \citet{Garn2008}. Given the uncertainty in uGMRT beam model (primary beam) and ionospheric fluctuations at low-frequency, it is essential to cross check with previous radio catalogs. The multi-frequency information available in literature for this field allow us to quantify any systematic offsets  in flux density and position of sources.  Counterpart of our catalog sources are identified using a 5$\arcsec$ search radius for all previous catalogs. Individual catalog has a flux density limit ($S_{\mathrm{limit}}$) based on completeness and sensitivity of that particular observation. We have scaled that flux density limit of different catalogs to 400 MHz  using a spectral index of $\alpha$ = -0.8 ($S_{\nu}\propto \nu^{\alpha}$). Hence, different flux density limits correspond to a flux cut at 400 MHz ($S_{\mathrm{cut,400MHz}}$). For comparison with other catalogs, we have  used only those sources whose flux density at 400 MHz is greater than the flux cutoff . Details of different survey parameters are mentioned in Table \ref{spectral_table}.  

\begin{table} 
\begin{center}
\caption{Details of previous radio catalogs. The frequency of observation, resolution, flux limit of a survey and flux cut at 400 MHz corresponding to flux limit (assuming $\alpha = -0.8$) are mentioned in different columns.  }
\label{spectral_table}
\begin{tabular}[\columnwidth]{lcccc}
\hline
\hline
Catalog & Frequency  & Resolution & $S_{\mathrm{limit}}^{\dagger}$ &  $S_{\mathrm{cut,400MHz}}$ \\
       &     & (arcsec) & (mJy)& (mJy)\\
 \hline
uGMRT & 400 MHz & 4.6$\arcsec$ & 0.10 & 0.10 \\
\hline

NVSS & 1.4 GHz  & 45$\arcsec$ & 2.5 & 6.8\\
\hline
FIRST  & 1.4 GHz & 5.4$\arcsec$ & 1.0 & 2.7\\
\hline
GMRT & 610 MHz  & 6$\arcsec$ & 0.27 & 0.37\\
\hline
GMRT & 325 MHz  & 9$\arcsec$ & 0.26 & 0.22\\
\hline
\hline
\end{tabular}
\end{center}
\flushleft{${\dagger}$ $S_{\mathrm{limit}}$ is the flux density limit of the corresponding catalog. \\
 }
\end{table}

\subsection{Flux density offset}
\label{Flux_offset}
  Uncertainties in the flux density scale (e.g \citealt{Perley2013}) and uGMRT beam (primary beam) model can cause for  systematic offsets in flux density. We have  compared uGMRT flux densities  with GMRT observation of ELAIS N1 at 610 MHz by \citet{Garn2008}. We follow the same selection criteria as described in \citet{Williams2016}.  We  have selected only high signal-to-noise  $\Big(S_{\mathrm{peak}} > 10\sigma\Big )$ sources in both maps. The minimum distance between any two sources in our map is restricted to be greater than 12$\arcsec$ (i.e. twice the size of the GMRT PSF at 610 MHz) to ensure that they are well isolated. We choose only a sample of compact sources (measured size less than the resolution at 610 MHz). The flux limit of the 610 MHz GMRT survey is 0.27 mJy \citep{Garn2008}, which corresponds to 0.37 mJy at 400 MHz assuming a spectral index -0.8. We have selected sources with flux density greater than this flux limit at 400 MHz (>0.37 mJy).  These restrictions finally gave us a sample of 122 sources for further analysis.    

We have multiplied the 610 MHz GMRT  flux density (originally in \citealt{Baars1977} scale) by 0.94 to place them on \citet{Perley2013} scale and scaled it to 400 MHz assuming a spectral index of $\alpha$ = -0.8.

\begin{figure}
\centering
\includegraphics[width=\columnwidth,height=2.5in]{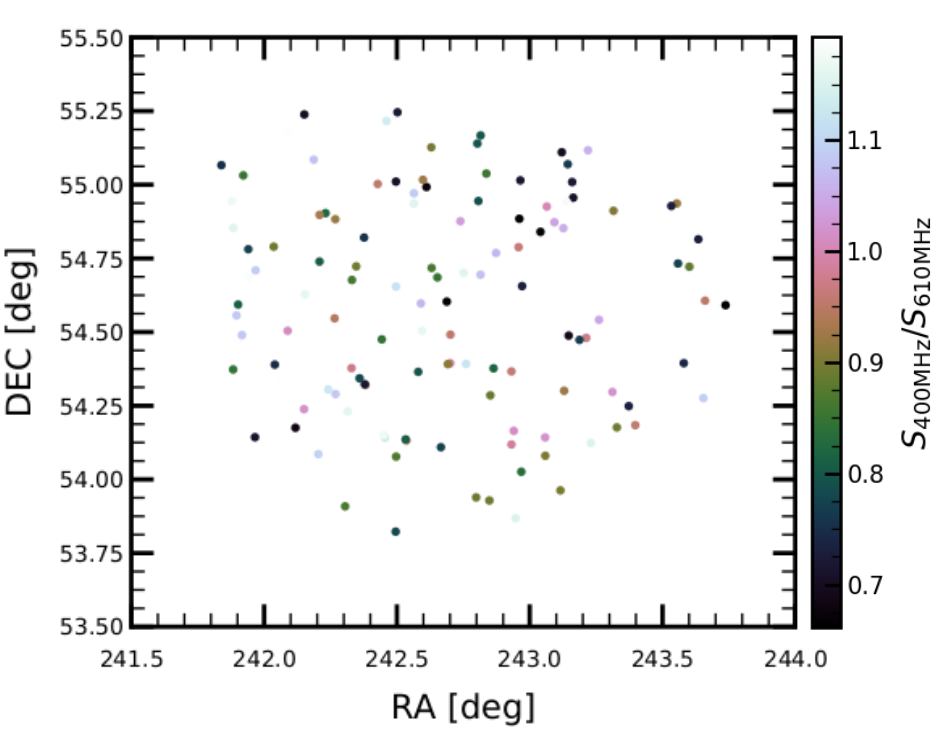} 
\caption{ Map of the ratios of integrated flux densities for high signal-to-noise, compact and isolated uGMRT 400 MHz sources with respect to GMRT 610 MHz sources. The colorscale is showing the flux density ratio.}
\label{flux_ratio}
\end{figure}

For this sample of sources, we have calculated the ratio of integrated flux densities between 400 MHz uGMRT and 610 MHz GMRT samples  $\Big(S_{\mathrm{400MHz}}/S_{\mathrm{610MHz}}\Big)$. We have found that the median of the ratio to be $1.01^{+0.3}_{-0.2}$ with errors from 16th and 84th percentiles. We have plotted RA and DEC of these sources  in the  Fig. \ref{flux_ratio}, where the colorbar shows the flux density ratio. We have not found any significant variation across the image.  In the central part of the map, the ratio is close to one for a significant number of sources.

To cross validate this result with other radio observations, we have performed a similar comparison with NVSS  \citep{Condon1998} , FIRST catalog \citep{Becker1995} and also with 325 MHz GMRT observation \citep{Sirothia2009} . Here again we restrict our requirements of  sample selection as described above.   We have determined the flux density ratio  for these samples after proper scaling. The median value of flux ratios together with the errors are  $0.95^{+0.2}_{-0.3}$, $1.06^{+0.3}_{-0.4}$, $1.09^{+0.2}_{-0.3}$  for NVSS, FIRST and 325 MHz GMRT catalog respectively.  

\begin{figure}

\includegraphics[width=\columnwidth]{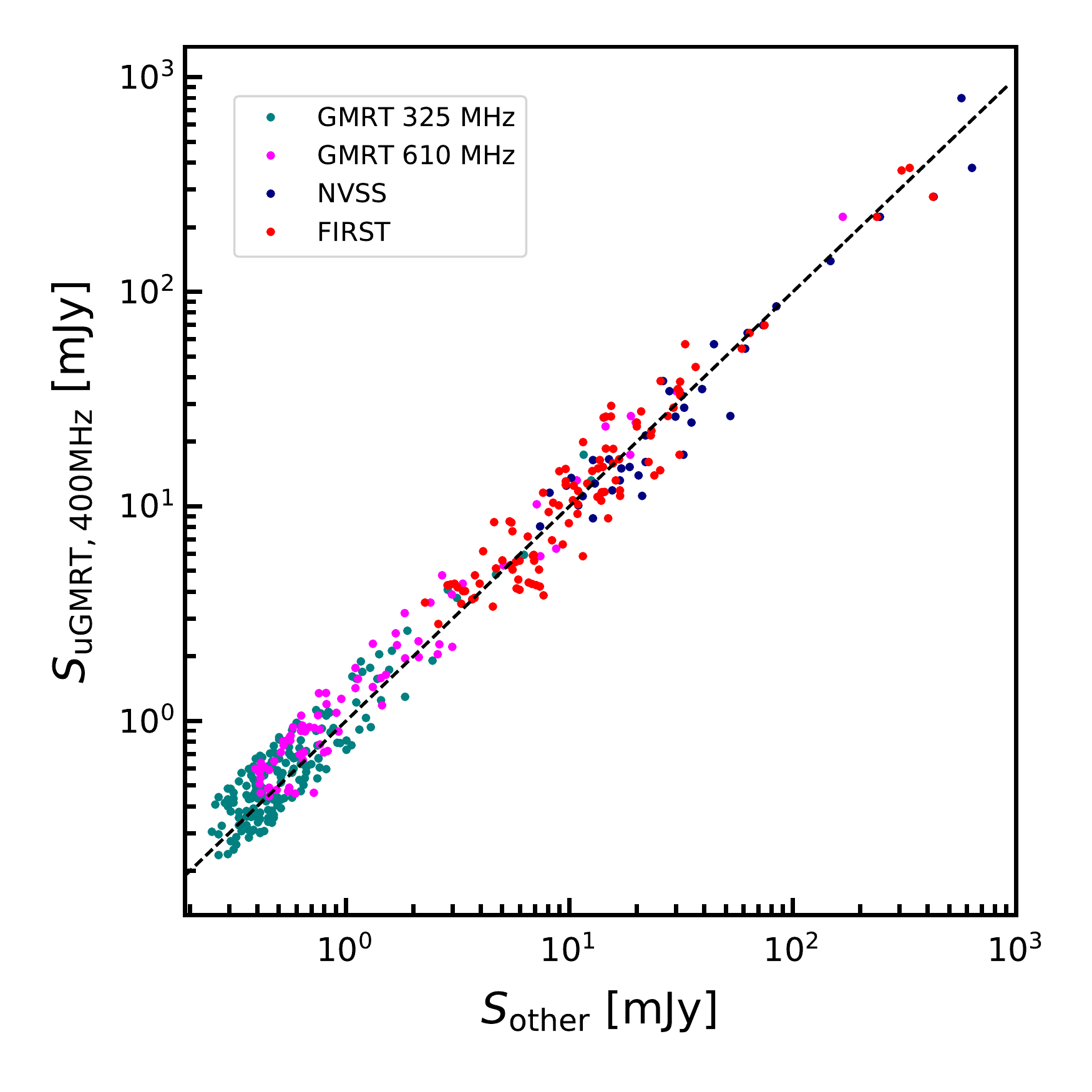}

\caption{ Comparison of total flux density of compact  sources measured at 400 MHz using uGMRT with other radio catalogs at different frequencies 325 MHz GMRT (green), 610 MHz GMRT (magenta),  NVSS (blue), FIRST (red). The black dashed line corresponds to $S_{\mathrm{uGMRT}}$/ $S_{\mathrm{other}}$ = 1.   }
\label{flux_ratio_hist}
\end{figure}
In Fig. \ref{flux_ratio_hist}, we have shown the flux densities of selected  sources at 400 MHz uGMRT observation as a function of flux densities of counterparts  in other catalog.  We have not found any significant deviation in flux density ratios for these different catalog comparisons. The median value is also close to 1 for all cases. We can say that the systematic effect is negligible here and we opt for no correction in the flux density due to systematic offsets.   

\subsubsection{Flux scale accuracy}

To check the overall reliability of the flux scale and to account for the uncertainties in spectral index, we have compared a small number of sources which are detected at higher frequency (FIRST 1.4 GHz) as well as in lower frequency (uGMRT 325 MHz;  \citealt{Sirothia2009}) maps. Here again we restrict our choice to compact , high signal-to-noise  and well isolated sources. We have properly scaled the flux densities to  \citet{Perley2013}  to put them in the same flux scale. Then, for this sample of  sources, we have first calculated the spectral index by comparing FIRST and 325 MHz flux densities and then predicted the uGMRT flux density at 400 MHz. The median value of spectral index for these sources is $\alpha$ = -0.77. 
\begin{figure}
\centering
\includegraphics[width=\columnwidth]{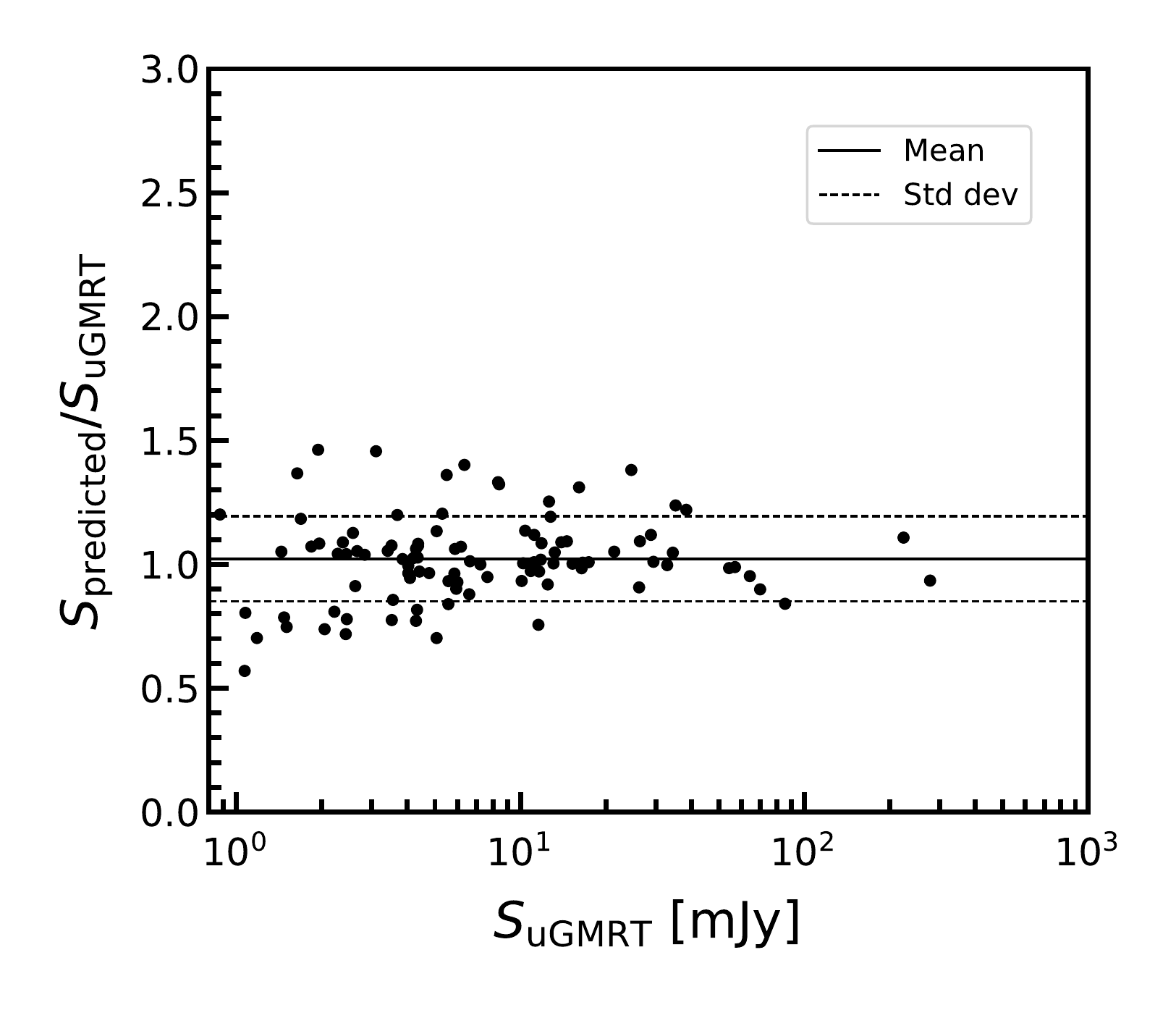}
\caption{Comparison between  flux densities measured at 400 MHz (uGMRT) and predicted flux densities using  325 MHz \citep{Sirothia2009} and FIRST catalog. The mean value of the ratio,  ($S_{\mathrm{predicted}}/S_{\mathrm{uGMRT}}$),  is $1.02\pm 017$.}
\label{prediction_flux}
\end{figure}
In Fig. \ref{prediction_flux}, we have shown predicted to measured flux density ratio as a function of uGMRT flux densities.  The mean flux density ratio is 1.02  with standard deviation of 0.17. We can conclude that the corrected uGMRT flux density is consistent with \citet{Perley2013} scale.

\subsection{Positional accuracy}

We have not done any direction-dependent calibration for this wide bandwidth uGMRT observation of the ELAIS N1 field. Phase only self-calibration can reduce fluctuation in phase  but only near the apparent bright sources.  There are  residual phase errors after final calibration procedure, causing   uncertainty in the source  positions. Ionospheric fluctuation at low-frequency also induces positional offsets.  Here we have assessed the astrometric accuracy of our catalog by comparing source positions with 1.4 GHz FIRST catalog \citep{Becker1995,Thyagarajan2011}. Due to high frequency, FIRST catalog has better resolution (5.4$\arcsec$)  and also ionospheric fluctuation is comparatively small. The positional accuracy of FIRST catalog is better than 1$\arcsec$ \citep{Becker1995}.


\begin{figure}
    \centering
    \includegraphics[width=\columnwidth,height=3.0in]{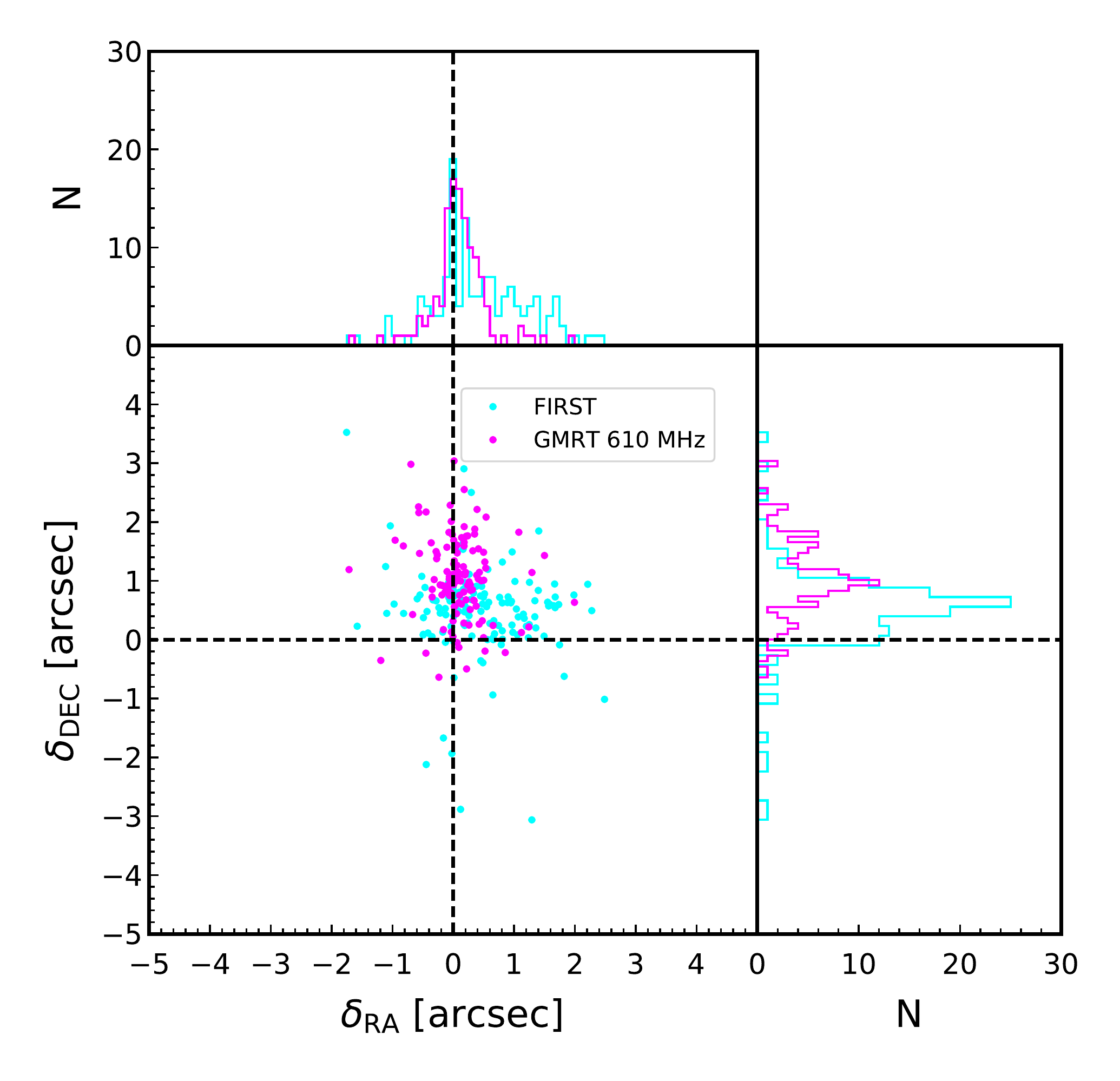}
    \caption{The offset in RA and DEC of the uGMRT sources at 400 MHz compared to FIRST catalog and GMRT 610 MHz observation.}
    \label{offset}
\end{figure}

Again we have selected a sample of small, isolated and compact sources following the criteria described in Sec. \ref{Flux_offset}.  This gives us a sample consists of 135 sources for comparison. We have  measured the offset in right ascension (RA) and declination (DEC) of these sources as (following \citealt{Williams2016}) : 
\begin{eqnarray}
\delta_{\mathrm{RA}} &=& \mathrm{RA}_{\mathrm{uGMRT}} - \mathrm{RA}_{\mathrm{FIRST}} \\  
\delta_{\mathrm{DEC}} &=& \mathrm{DEC}_{\mathrm{uGMRT}} - \mathrm{DEC}_{\mathrm{FIRST}} \nonumber
\end{eqnarray}

The median offset in RA and DEC are 0.28$\arcsec$ and 0.56$\arcsec$.  There is no systematic  variation of positional offset ($\delta_{\mathrm{RA}}$ and $\delta_{\mathrm{DEC}}$)  across the FoV. We have done similar analysis with GMRT 610 MHz catalog \citep{Garn2008}. Here the mean offset in RA and DEC are 0.06$\arcsec$  and 1$\arcsec$. The offset in DEC is  slightly higher in this case. But ionosphere is more unstable at 610 MHz  GMRT observation and also no direction-dependent calibration has been performed for this observation \citep{Garn2008}. In Fig. \ref{offset}, the histogram of offset in RA and DEC for both catalogs is shown.  Given the  pixel size of 1.5$\arcsec$ of uGMRT image, these offsets are negligible. 

We made a correction in the final catalog for uGMRT source positions  with a constant offset, i.e,  $\delta_{\mathrm{RA}}$ = 0.28$\arcsec$ and $\delta_{\mathrm{DEC}}$ = 0.56$\arcsec$ (based on FIRST catalog offsets).

\subsection{Spectral index distribution}
Characterization of spectral properties of sources in ELAIS N1 field is done by comparing flux densities with  previous high frequency radio catalogs. For comparison, we have used  610 MHz GMRT observation of the same field \citep{Garn2008} and  FIRST (1.4 GHz) and NVSS (1.4 GHz) catalogs.  We follow the same source selection procedure as in Sec \ref{Flux_offset}. The sample includes compact, isolated and high SNR sources, whose flux density values are above the flux limit of corresponding catalogs. The number of sources used to estimate spectral index distribution for different catalog comparison are: 44 (NVSS), 135 (FIRST) and 80 (GMRT, 610 MHz).   

\begin{figure}
\centering
\includegraphics[width=\columnwidth]{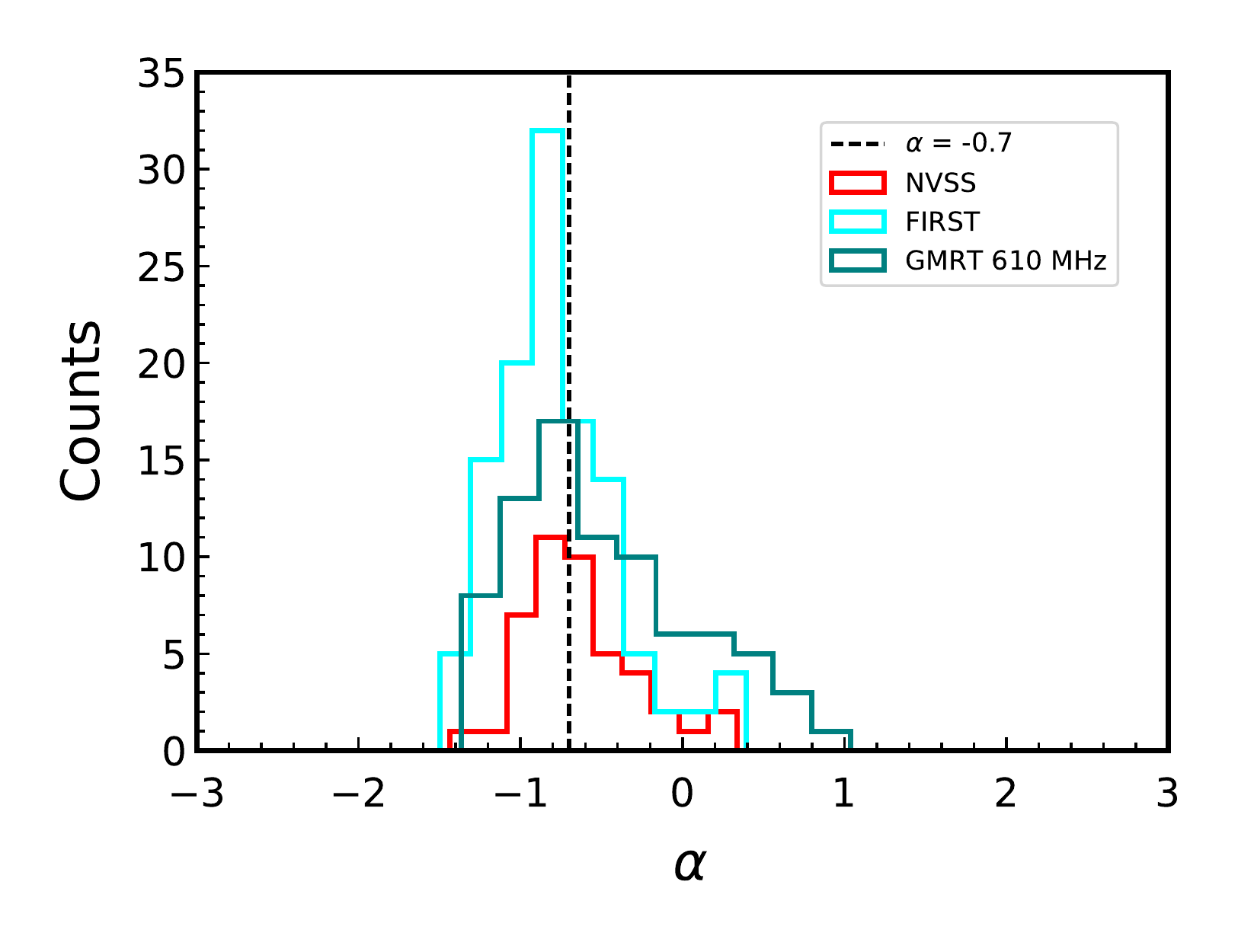}
\caption{ The histogram of measured spectral indices of sources in this field after matching with different  catalogs using a 5$\arcsec$ match radius. The black  dashed line corresponds to   $\alpha$ = -0.7.  } 
\label{sp}
\end{figure}

 We have assumed a synchrotron power-law distribution with single spectral index, i.e, $S_{\nu} \propto \nu^{\alpha}$, where $\alpha$ is the spectral index.  We have compared the flux density of matched sources between two catalogs  and then estimate the $\alpha$ value.  In Fig. \ref{sp}, we have shown the histogram of $\alpha$ for the sources in our catalog matched to other three different catalogs. The median spectral indices with errors from  16th  and 84th percentile for different catalogs are : $-0.81^{+0.28}_{-0.32}$ (1.4 GHz FIRST),  $-0.70^{+0.31}_{-0.24}$ (1.4 GHz NVSS) and $-0.68^{+0.36}_{-0.52}$ (610 MHz GMRT). \citet{Garn2008} have reported a spectral index value of -0.7 by comparing flux densities with FIRST catalog. \citet{Sirothia2009} measured a median spectral index of -0.83 after matching sources with 1.4 GHz FIRST catalog. They have also reported a more steeper median value of spectral index -1.28 after comparing flux densities with 610 MHz GMRT catalog of \citet{Garn2008}. Here, the median value of $\alpha$ estimated after comparing with  different radio catalogs is close to -0.7, which is in agreement  with previous measurements. A detailed study of spectral index of sources using multi-frequency data as well as analysis of in band uGMRT spectral indices is deferred to future work.
\section{Source counts}
\label{sec.source_counts}
We have estimated the differential source counts based on wide-band flux densities  from P{\tiny Y}BDSF catalog output. At low-frequency, distribution of sources as a function of flux density is important to understand population of different radio galaxies. We know from simulation \citep{Wilman2008} as well as from different observations that star forming galaxies (SFGs) and the radio quiet quasars (RQQ) are most dominant populations at faint flux densities. But, there are very few observational constraints on source population at sub mJy level, mainly below 0.5 mJy. Characterization of the spatial and spectral nature of the foreground sources down to $\mu$Jy level flux density is crucial for telescopes like LOFAR, MWA, HERA and SKA in order to detect the faint cosmological HI 21 cm signal.

Here, we have measured the differential source counts at 400 MHz down to 120 $\mu$Jy (> 8$\sigma$). But, direct quantification of source counts based on P{\tiny Y}BDSF output does not contemplate true extragalactic source distribution, specially at low frequencies and at faint end of flux density bins. We need to correct for incompleteness, false detection rate, Eddington bias, resolution biases as well as visibility area effects. These correction factors are described in detail below.

\subsection{False detection rate}
False detection rate defines as the number of spurious sources detected by the source finding package (P{\tiny Y}BDSF) as real ones due to noise spikes and artifacts. 
\begin{figure}
    \centering
    \includegraphics[width=\columnwidth]{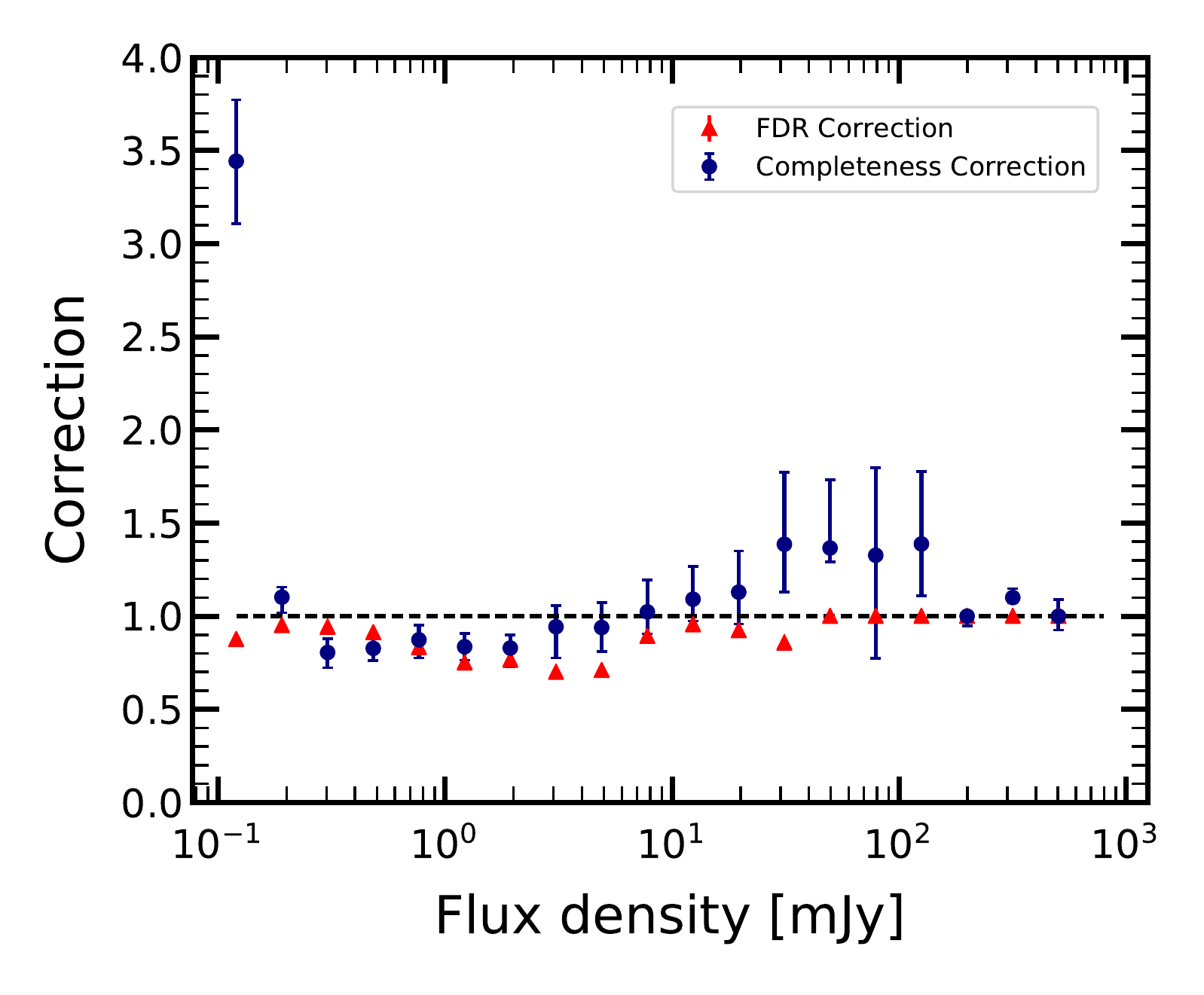}
    \caption{Correction factors due to false detection rate or FDR (red) and completeness (green) as a function of integrated flux density}
    \label{FDR}
\end{figure}
If the distribution of noise in the image is symmetric about zero, i.e, positive noise spikes have equivalent negative spikes in the image, then number of falsely detected (spurious) sources would be identical to the number of ``negative'' sources in the inverted (or, negative) image. In order to quantify this, we run P{\tiny Y}BDSF on the inverted (negative) image with exactly same parameters as used in our source finding algorithm (Sec. \ref{sec.source_catalog}). We have detected  a total of 243 sources with negative peaks less than -5$\sigma$.

To correct the flux density bins for FDR, we have binned the number of negative sources detected in the inverted image in 20 logarithmic bins and compared this with the positive sources detected in the original image. Fraction of real sources in each bin is defined as \citep{Hale2019} - 
\begin{equation}
     \textit{f}_{\mathrm{real},\textit{i}} = \frac{\textit{N}_{\mathrm{catalog},\textit{i}} - \textit{N}_{\mathrm{inv},\textit{i}}}{\textit{N}_{\mathrm{catalog},\textit{i}}},
    \label{FDR_eqn}
\end{equation}

where  $\textit{N}_{\mathrm{inv},\textit{i}}$ and $\textit{N}_{\mathrm{catalog},\textit{i}}$ are the number of  detected sources in $i^{th}$ flux density bin for inverted and original image respectively. The correction factor due to false detection  is shown in Fig. \ref{FDR}. The errors in FDR are calculated using Poissonian errors.  We have multiplied this fraction (Eqn. \ref{FDR_eqn})  to the number of sources detected in each flux density bin in the original catalog.

\begin{table*}
\caption{Euclidian-normalized differential source counts for ELAIS N1 field.}
\label{table_source_count}
\scalebox{1.0}{
\begin{tabular}{l l l l l l l l}
\hline
\hline
 S & $S_{c}$  & N   & $ S^{2.5}$dN/dS & FDR & Completeness &  Corrected $ S^{2.5}$dN/dS  \\
(mJy) & (mJy) &       &  ($\mathrm{Jy}^{1.5}$sr$^{-1}$)  &       &    & ($\mathrm{Jy}^{1.5}$sr$^{-1}$)           \\
 \hline
 0.120-0.191 &  0.166   & 218   & $6.635 \pm 0.183$ & $0.876 \pm 0.004$ & $3.44^{+0.32}_{-0.33}$ & $ 20.01 \pm 0.55$   \\
  & & & & & &  \\
 0.191-0.303 &  0.244   & 688  & $18.269 \pm 0.391$ & $0.951 \pm 0.001$ & $1.10^{+0.05}_{-0.08}$ & $19.14 \pm 0.41$  \\
   & & & & & &  \\
 0.303-0.482 &  0.380   & 701  & $27.627 \pm 0.665$  & $0.942 \pm 0.001$ & $0.80^{+0.07}_{-0.08}$ & $ 20.95 \pm 0.50$  \\
   & & & & & &  \\
 0.482-0.766 &  0.603   & 381  & $27.542 \pm 0.938$ & $0.913 \pm 0.002$ & $0.82^{+0.07}_{-0.06}$ & $20.79 \pm 0.71$   \\
   & & & & & &  \\
 0.766-1.218 &  0.943   & 191  & $25.957 \pm 1.263$ & $0.832 \pm 0.006$ & $0.87^{+0.07}_{-0.09}$ & $18.85 \pm 0.92 $   \\
   & & & & & &  \\
 1.218-1
 935&  1.511   & 112  & $30.882 \pm 1.971$ & $0.750 \pm 0.011$ & $0.83^{+0.07}_{-0.07}$ & $19.36 \pm 1.24$   \\
   & & & & & &  \\
 1.935-3.076 &  2.344   & 68  & $35.184 \pm 2.887 $ & $0.765 \pm 0.013$ & $0.82^{+0.07}_{-0.10}$ & $22.31 \pm 1.83$   \\
   & & & & & &  \\
 3.076-4.889 &  3.956   & 50   & $60.132 \pm 5.760 $ & $0.700 \pm 0.019$ & $0.94^{+0.11}_{-0.17}$ & $39.69 \pm 3.80$   \\
   & & & & & &  \\
 4.889-7.772 &  5.850   & 31   & $62.327 \pm 7.585 $ & $0.710 \pm 0.024$ & $0.93^{+0.13}_{-0.12}$ & $41.55 \pm 0.85$   \\
   & & & & & &  \\
 7.772 -12.353 &  10.058   & 28   & $137.190 \pm 17.573 $ & $0.893 \pm 0.010$ & $1.02^{+0.17}_{-0.12}$ & $125.32 \pm 16.05$   \\
   & & & & & &  \\
12.353-19.635 &  15.092   & 22   & $186.968 \pm 27.023 $ & $0.955 \pm 0.005$ & $1.09^{+0.17}_{-0.11}$ &  $194.80 \pm 28.15 $   \\
   & & & & & &  \\
19.635-31.209 &  24.877   & 13   & $242.412 \pm 45.584 $ & $0.923 \pm 0.010$ & $1.12^{+0.22}_{-0.17}$ & $252.61 \pm 47.50$   \\
   & & & & & &  \\
31.209 - 49.607 &  37.404   & 7   & $227.605 \pm 58.330$ & $0.857 \pm 0.026$ & $1.38^{+0.38}_{-0.25}$ & $ 270.35 \pm 69.28$   \\
   & & & & & &  \\
49.607-78.849 &  61.337  & 4   & $281.753 \pm 95.524$ & $1.000$ &   $1.36^{+0.36}_{-0.07}$ & $ 384.87 \pm 130.48$ \\
   & & & & & &  \\
 78.849 - 125.330 & 85.496 & 1   & $101.645 \pm 68.924$ & $1.000$ & $1.32^{+0.47}_{-0.55}$ & $ 134.88 \pm 91.46$ \\
   & & & & & &  \\
 125.330 -199.212 & 138.951 & 1   & $215.329 \pm 146.014$ & $1.000$  & $1.38^{+0.38}_{-0.27}$ & $ 298.87 \pm 202.66$ \\
   & & & & & &  \\
 199.212 - 316.645 & 250.260 & 2    & $1179.466 \pm 565.546$   & $1.000$ & $1.00^{+0.01}_{-0.05}$ & $ 1179.46 \pm 565.54$ \\
   & & & & & &  \\
 316.645 - 503.305 & 372.677 & 2   & $2008.035 \pm 962.850$ & $1.000$ & $1.00^{+0.05}_{-0.02}$ & $ 2208.83 \pm 1059.13$ \\
   & & & & & &  \\
 503.305 - 800.0 & 798.478 & 1    & $4244.269 \pm 2878.116$ & $1.000$ & $1.00^{+0.08}_{-0.07}$ &  $ 4244.26 \pm  2878.12$ \\
   & & & & & &  \\
 \hline 
\end{tabular}}
\flushleft{Notes: This table includes the flux density bins, central of flux density bin, the raw counts, normalized source counts, False Detection Rate (FDR), completeness and corrected normalized source counts.}
\end{table*}
\subsection{Completeness}
A source catalog constructed using  P{\tiny Y}BDSF output is not complete. There are some factors which can cause  for underestimation as well as over-estimation of  the source counts. This makes the catalog incomplete. To quantify those, we carried out simulation in the image plane. Incompleteness means given a flux density limit, we are still unable to detect sources above that limit due to varying noise in the image. This results in underestimation of source counts near the flux density detection limit.  Eddington bias causes noise to redistribute low flux density sources in higher fluxes. Due to steep source counts at low flux density bins, this bias is significant near the detection limit. As a consequence, there may be  boost in source counts in the faintest bins. Magnitude of this boost is governed by signal-to-noise ratio and source count slope.

 Resolution bias signifies that the detection probability of a resolved source is less than  point-like sources  in  our peak flux density selection during P{\tiny Y}BDSF run. For a extended source, the peak flux density may be significantly reduced that it can not be detected above the noise. Although these extended sources have same integrated flux density as the unresolved ones, we are unable to detect them and hence resolution bias reduces our source counts. 

We have quantified these biases by  injecting 1000 sources into our primary beam  corrected image (not the residual rms map as in \citealt{Williams2016}).  Out of these, 100 sources (10\%) are extended, i.e., sizes are greater than beam size. We scatter the sources randomly in the image plane.  The flux densities of simulated sources are drawn randomly from a power law distribution (dN/dS $\propto$ $S^{-1.6}$; \citealt{Intema2011,Williams2013}) between 80 $\mu$Jy to 1 Jy. We have created 100 such simulations. These simulations inherently take into account the confusion of sources and  visibility area of sources at different flux density bins \citep{Hale2019,Franzen2019,Williams2016}. 
\begin{figure*}
\centering
\includegraphics[width=5in]{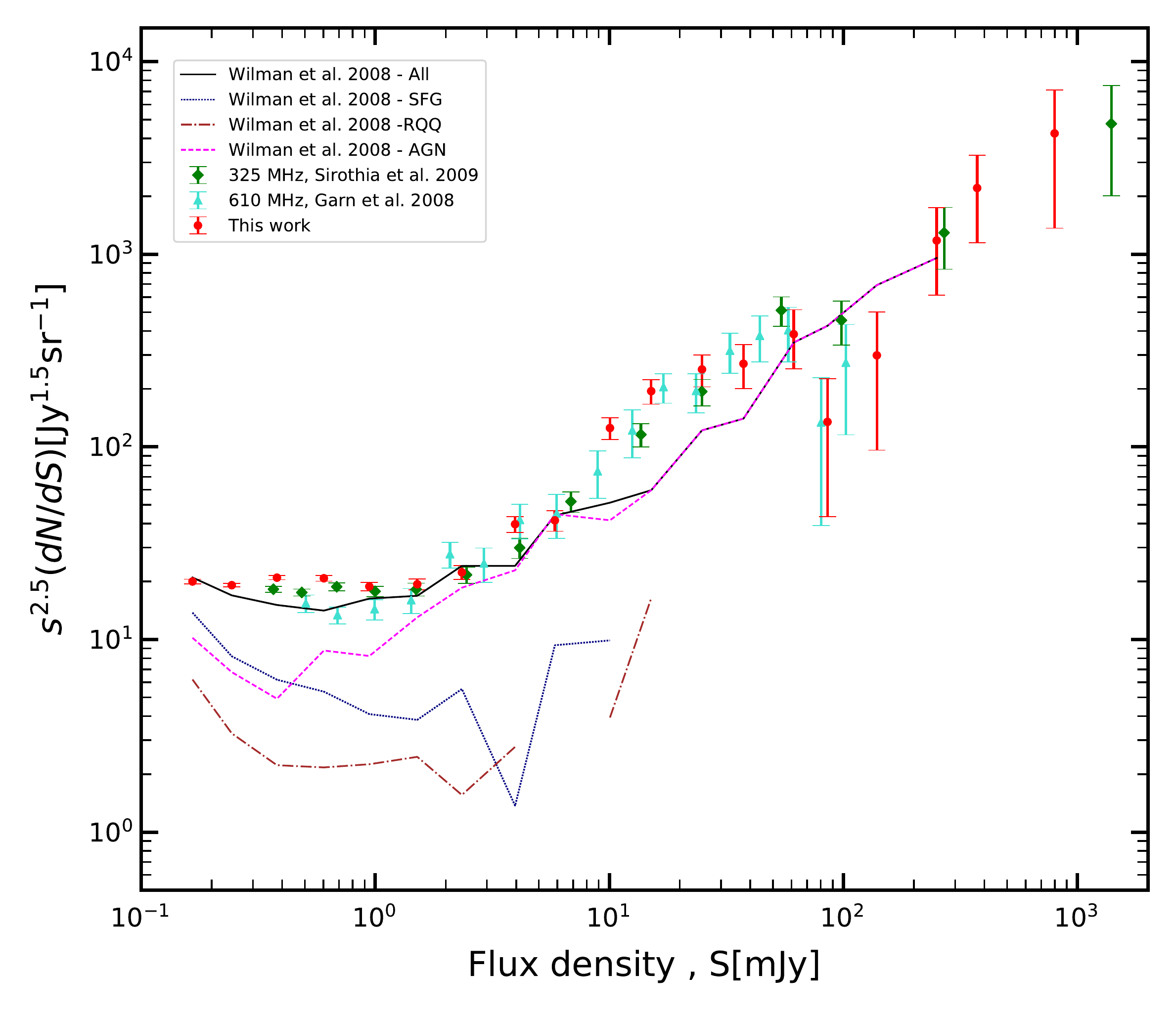}
\caption{Euclidian-normalized differential source counts for the uGMRT 400 MHz observation of ELAIS N1 field. The red circles show the observed source counts after correction factors have been applied. For comparison, we also plot 325 MHz (green) \citep{Sirothia2009}, 610 MHz (cyan) \citep{Garn2008} source counts of same field after scaling to 400 MHz (using $\alpha = -0.8$). We have also shown model source counts from  the $S^{3}$ simulation : all sources (black line), SFG (blue), RQQ (brown) and AGN (magenta). }
\label{dn_ds}
\end{figure*}
For each simulated image, we have extracted sources using P{\tiny Y}BDSF following the same criteria as described in Sec. \ref{sec.source_catalog}. There were sources prior to our simulation in the original image.  We already have the source   catalog corresponding to the original image. Now after detection of  sources from the simulated image, we have subtracted the original sources  from the post-simulation source counts (which consists of  injected sources and original sources). We have  binned these sources  in 20 logarithmic bins in flux density.  The correction factor then calculated as  (following \citealt{Hale2019}) - 
\begin{equation}
     \mathrm{Correction}_{,\textit{i}} = \frac{\textit{N}_{\mathrm{injected},\textit{i}}}{\textit{N}_{\mathrm{recovered},\textit{i}}}
\end{equation}
here, $\mathrm{Correction}_{,\textit{i}}$ is the completeness correction factor in the $i^{th}$ flux density bin. $\textit{N}_{\mathrm{injected},\textit{i}}$ is the number of injected sources and  $ \textit{N}_{\mathrm{recovered},\textit{i}}$ is the number of sources recovered after subtracting original pre-simulation sources in the $i^{th}$ bin. This method of quantifying completeness  already takes into account the   resolution bias as well as the Eddington bias \citep{Hale2019}. The completeness correction factor is shown in Fig. \ref{FDR}. We are quoting the median value of 100 simulations for each flux density bins as a correction factor and the associated errors are from 16th and 84th percentiles.

\subsection{Differential Source count}

 We have estimated the Euclidian-normalized differential source counts from the source list generated by P{\tiny Y}BDSF. We have corrected the source counts for FDR and completeness. The correction factors are multiplicative  to the original source counts.  We have also corrected for effective area for different flux density bins over which a source can be detected. This is due to the fact that the noise is varying significantly across the image (see Fig. \ref{rms}). Hence, faint sources can not be detected over the full image. So, we have found out the fraction of area (f) over which a source with a given flux density  can be detected (its visibility area) and weighted the source counts by the reciprocal of that fraction \citep{Windhorst1985}.   The normalized source counts can be seen in Fig. \ref{dn_ds}. We have binned the sources in 20 logarithmic bins in flux density down to 120 $\mu$Jy. This is the deepest source counts at this low-frequency. The error in source count for each bin is  Poisson errors.  The source counts and associated errors are given in table \ref{table_source_count}. We have compared this source counts with 325 MHz \citep{Sirothia2009} and 610 MHz \citep{Garn2008} GMRT source counts of ELAIS N1 field after scaling to 400 MHz assuming a spectral index of -0.8.  These source counts are in agreement with our findings. 
 
 We have also compared our source counts with  $S^{3}$ -SKADS simulation by \citet{Wilman2008}. We have taken 1.4 GHz flux densities of $S^{3}$ simulation and scaled it to 400 MHz using $\alpha = -0.8 $.
 SKADS-simulation  uses different multi-frequency observation to model luminosity function, clustering of sources, classification of different sources, etc and gives a synthetic radio catalog (see \citealt{Wilman2008} and the references therein).   We have shown  the source counts of SFG, RQQ and AGN and all sources (combination of all) in Fig. \ref{dn_ds}. It is observed that the source population of RQQ and SFG's are increased at low flux densities and give rise to flattening in the total source counts below 1 mJy. We have also found a similar feature in the normalized source counts signifies the increased population of SFG and RQQ at low flux density bins. Our observed counts is consistent with this simulated model.  However, our observed counts is little higher than $S^{3}$ simulation in the flux density range 10 mJy to 100 mJy. The exact reason behind this excess is unknown. However,  the models used to generate the simulated catalog in SKADS are based on high frequency data available in literature \citep{Wilman2008,Williams2016}. So, some deviation may also be plausible.

 We have found that completeness correction is most dominant effect in low flux density bins whereas FDR correction is not large at these flux densities. Another possible error can be induced by incorrect primary beam model of uGMRT.    
 The primary beam pattern may change due to antenna movement in azimuth-elevation while tracking the target field across the sky. Also, there can be errors in the estimation of the primary beam pattern from relevant data. In order to understand the effect of these errors/deviations in the primary beam pattern, we have considered about 10\% error around the best-fitted values of the  four parameters of the primary beam model of uGMRT at Band-3 (see Eqn. \ref{PB_equation}). We have estimated the normalized dN/dS curve with the errors in the four parameter values. Our results show no significant deviation from the normalized dN/dS obtained with best-fitted values of the four beam parameters. Hence, we can conclude that this curve is robust against any beam errors within 10\%.

\section{Spectral Evolution of DGSE power spectra}
\label{sec.DGSE}
After removal of point sources  from the observed data set, DGSE is still higher than the HI signal by orders of magnitudes. The smooth spectral behavior of foregrounds  holds promise to extract the faint cosmological signal amidst these bright foregrounds. But extracting the signal requires knowledge of spatial as well as spectral nature of foregrounds. Here we have quantified how amplitude of angular power spectrum of DGSE is evolving as a function of frequency.

DGSE is generally modelled  as a power law in both   angular scale and frequency (see Eqn. \ref{power_law_model}).  This is an empirical model of foregrounds. Several previous observations have measured the APS of   DGSE for different fields and measured the value of the power law index ($\beta$) lies between [1.5 to 3.0] \citep{Ali2008,Iacobelli2013, Bernardi2009, Ghosh2012, Iacobelli2013,Choudhuri2017}.  
\citet{La Porta2008} has studied 408 MHz Haslam map \citep{Haslam1982} and 1420 MHz map created  by \citet{Reich1988} after combining Northern and Southern sky survey. They have measured  the APS of Galactic synchrotron emission ($\mathcal{C}_{\mathcal{\ell}}$) for different Galactic latitudes. Then using the mean APS at two frequencies (408 MHz and 1420 MHz) they have calculated mean spectral index ($\alpha$) using the relation  \citep{La Porta2008}:
\begin{equation}
  \big <\mathcal{C}_{\mathcal{\ell}}(\nu_{1})\big>
  =\big<\mathcal{C}_{\mathcal{\ell}}(\nu_{2})\big> \Big(\frac{\nu_{1}}{\nu_{2}}\Big)^{(-2\alpha)}
\end{equation}

(Note that: \citealt{La Porta2008} has used $\alpha$ as the power law index of APS and $\beta$ as the mean spectral index. So, our notation is just opposite to them)

\begin{figure*}
\centering
\begin{tabular}{c}

\includegraphics[width=6.25in,height=1.8in]{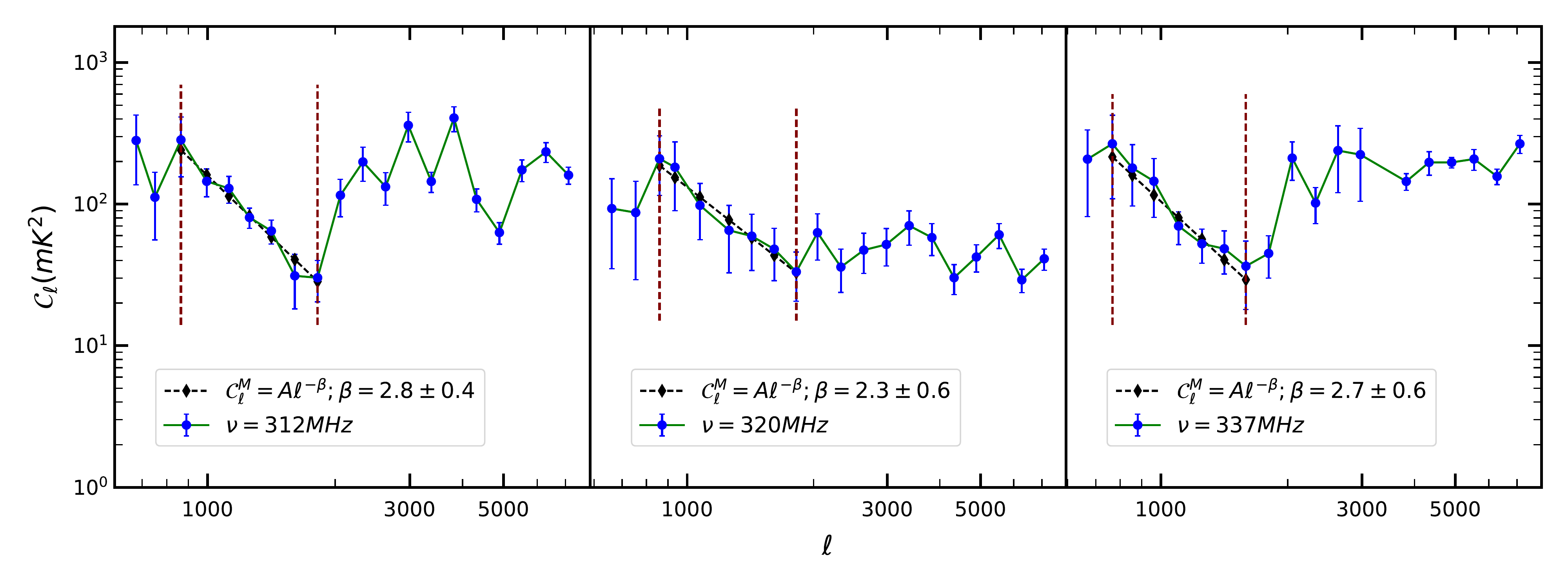}\\
\includegraphics[width=6.25in,height=1.8in]{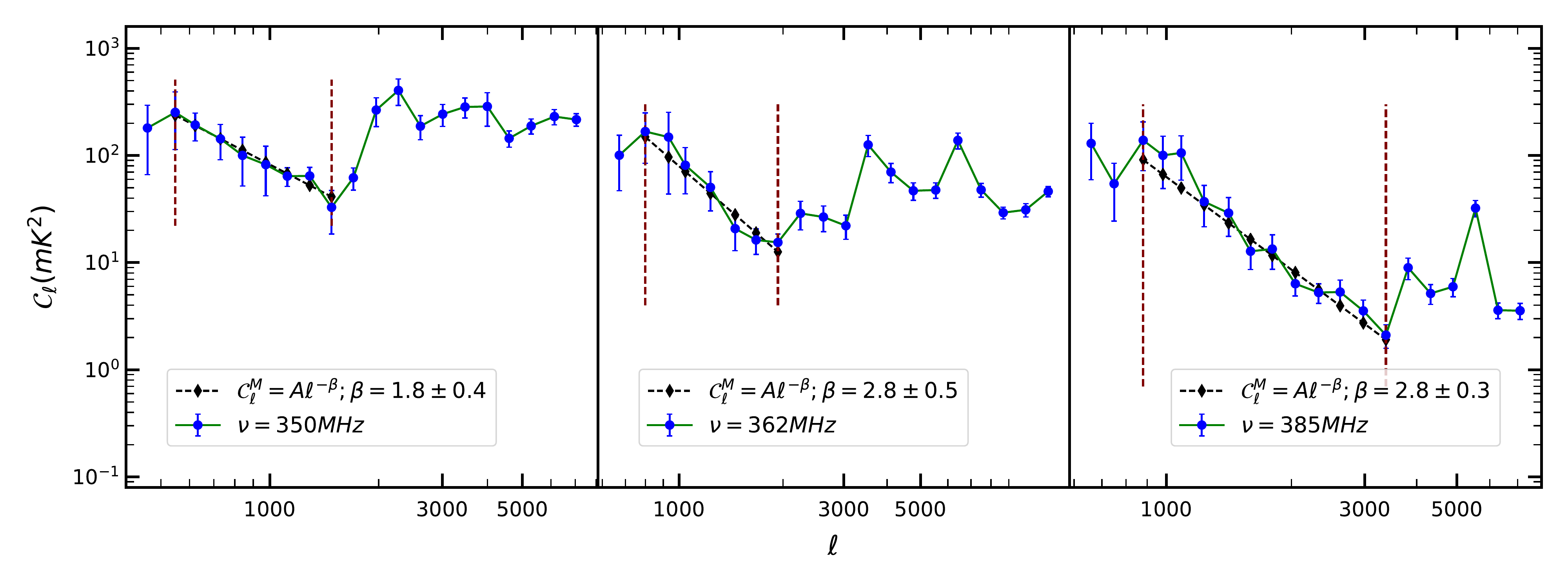}\\
\includegraphics[width=6.25in,height=1.8in]{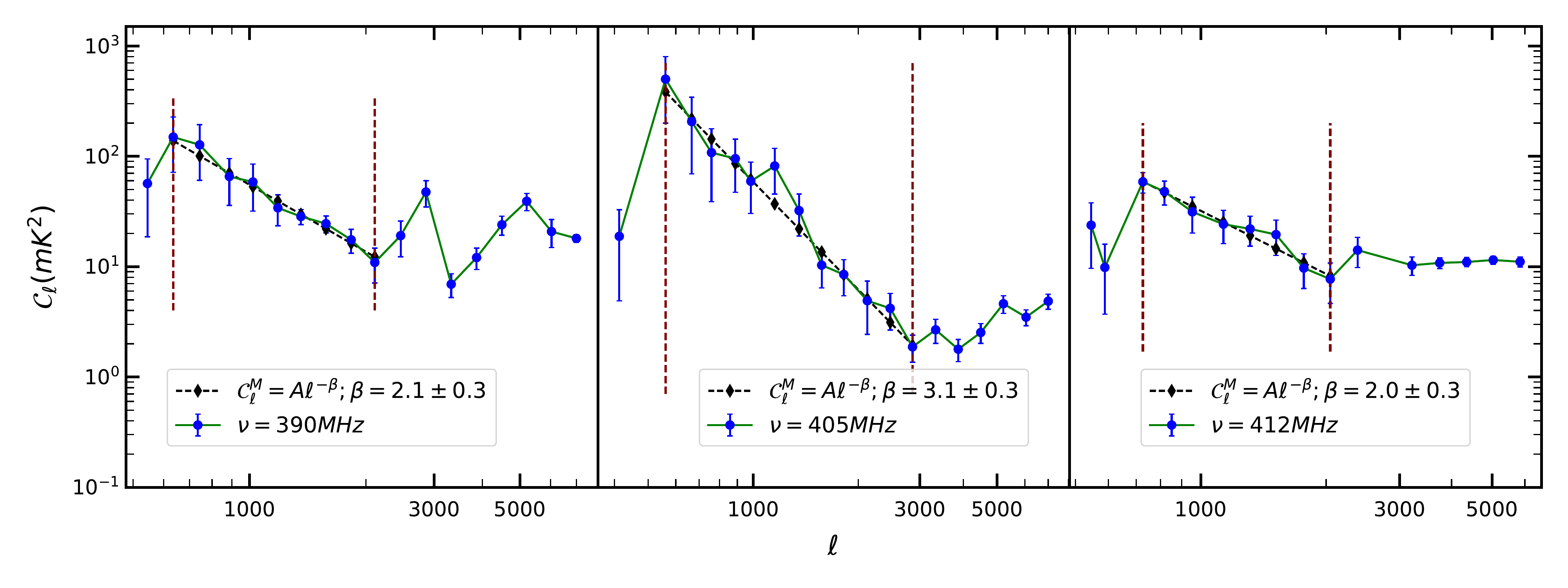}\\
\includegraphics[width=6.25in,height=1.8in]{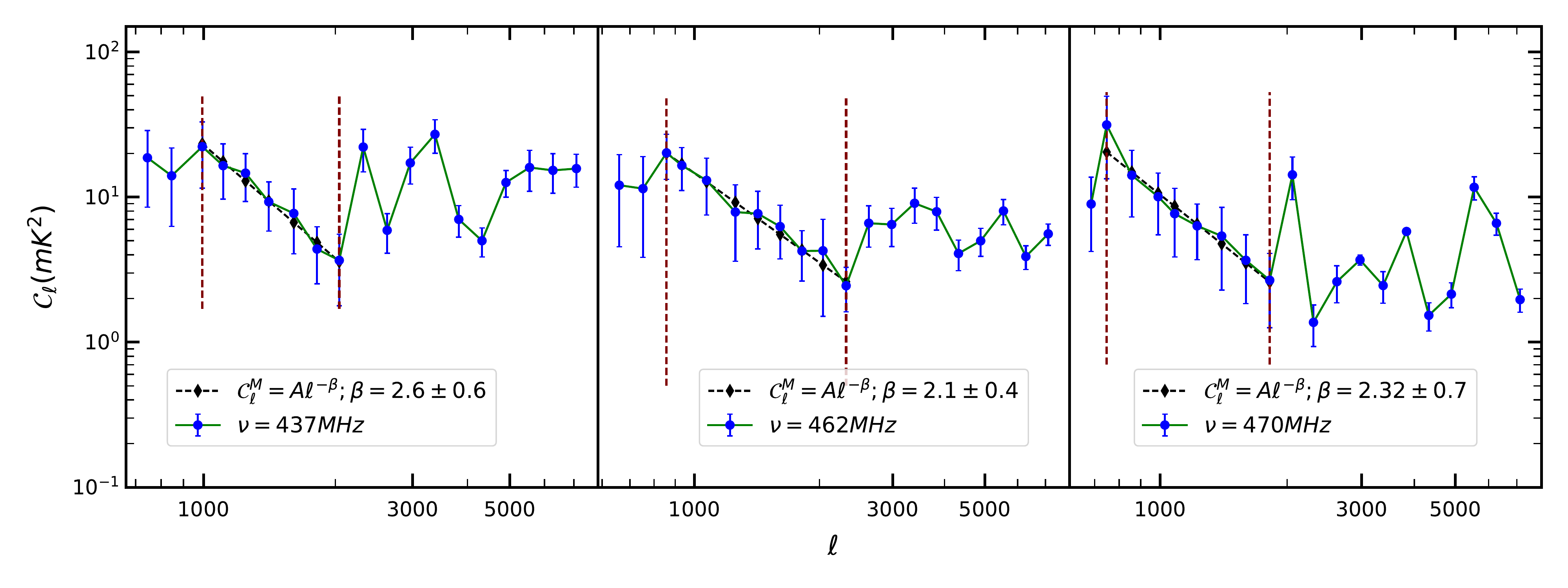}\\
\includegraphics[width=2.2in,height=1.8in]{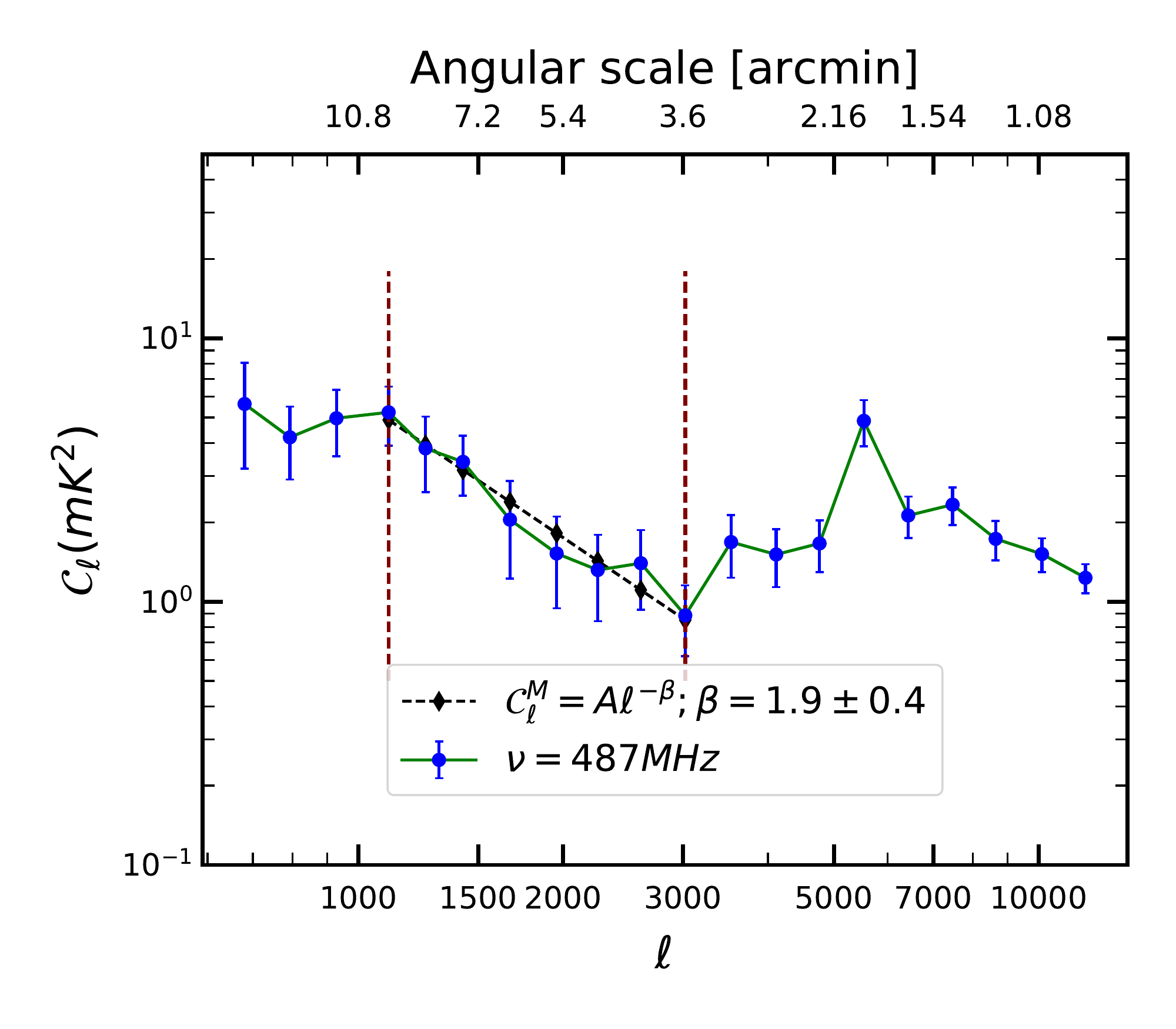}\\
\end{tabular}
\caption {The estimated angular power spectrum ($\mathcal{C}_{\mathcal{\ell}}$) with with $1-\sigma$ error bar (green curve) as a function of angular multipole $\mathcal{\ell}$ for 13 sub-bands. The vertical dashed lines (in maroon) shows $\mathcal{\ell}$ range to fit a power law model and the black dashed line shows the best-fitting, $\mathcal{C}_\ell^{M} = A \ell^{-\beta}$. The value of angular power law index $\beta$ is mentioned in each plot. In the last panel, we also show the angular scale  corresponding to the $\mathcal{\ell}$ range probed here.}
\label{power_plots}
\end{figure*}  

 The obtained  mean spectral index lies between [2.9 to 3.2] for different Galactic latitudes. Using this mean spectral index, they have extrapolated $\mathcal{C}_{\mathcal{\ell}}$ to 23 GHz and check the consistency of synchrotron APS with the WMAP observation of foregrounds at 23 GHz. They have derived the mean spectral index by comparing  amplitude of APS at  two frequencies and extrapolate this to higher frequency.  Different astrophysical components contribute in a different manner to the APS of foregrounds at different frequencies. As a consequence of that, fluctuation in DGSE can also vary as a function of frequency. Hence, estimating $\alpha$ based on only two discrete frequency samples may overlook the detail intricacies of synchrotron power spectrum as a function of frequency. 
    
We have used the  wide bandwidth (200 MHz) data of ELAIS N1 field  to find out the spectral behavior of fluctuation in  DGSE, i.e, spectral evolution of $\alpha$. First, we have subtracted the point source model (generated during CLEANINg) from the calibrated data set by using \textit{UVSUB} in {\tiny CASA}. The residual data (after subtraction) mainly consists of DGSE and  residual point sources below the noise level. Then, we have used the Tapered Gridded Estimator (TGE) \citep{Choudhuri2014,Choudhuri2016} to quantify the  APS of DGSE from the residual visibility data set. TGE  uses visibility correlation after gridding the calibrated data on a regular grid and subtracts the positive noise bias (by not including self-correlation of visibilities) to give unbiased estimate of the angular power spectrum ($C_{\mathcal{\ell}}$) (For more details please see: \citealt{Choudhuri2014,Choudhuri2016}).

 We have shown in \citet{Chakraborty2019} that with large tapering of FoV, we can estimate the angular power spectrum of diffuse radiation even without direction-dependent calibration. We can neglect the undesired effects of bright sources at large distance from the centre of the FoV in the estimation of $C_{\mathcal{\ell}}$, by using the higher tapering of sky response for direction-independent calibration in comparison with direction-dependent calibration \citep{Chakraborty2019}. Here, since we have not done any direction-dependent ionospheric calibration, we have used the same tapering parameter (f = 0.5) as used in \citet{Chakraborty2019} for direction-independent calibration. This ensures that the estimation of $C_{\mathcal{\ell}}$ is not be affected by direction-dependent calibration effects.
\begin{figure*}
    \centering
    \includegraphics[width=5in]{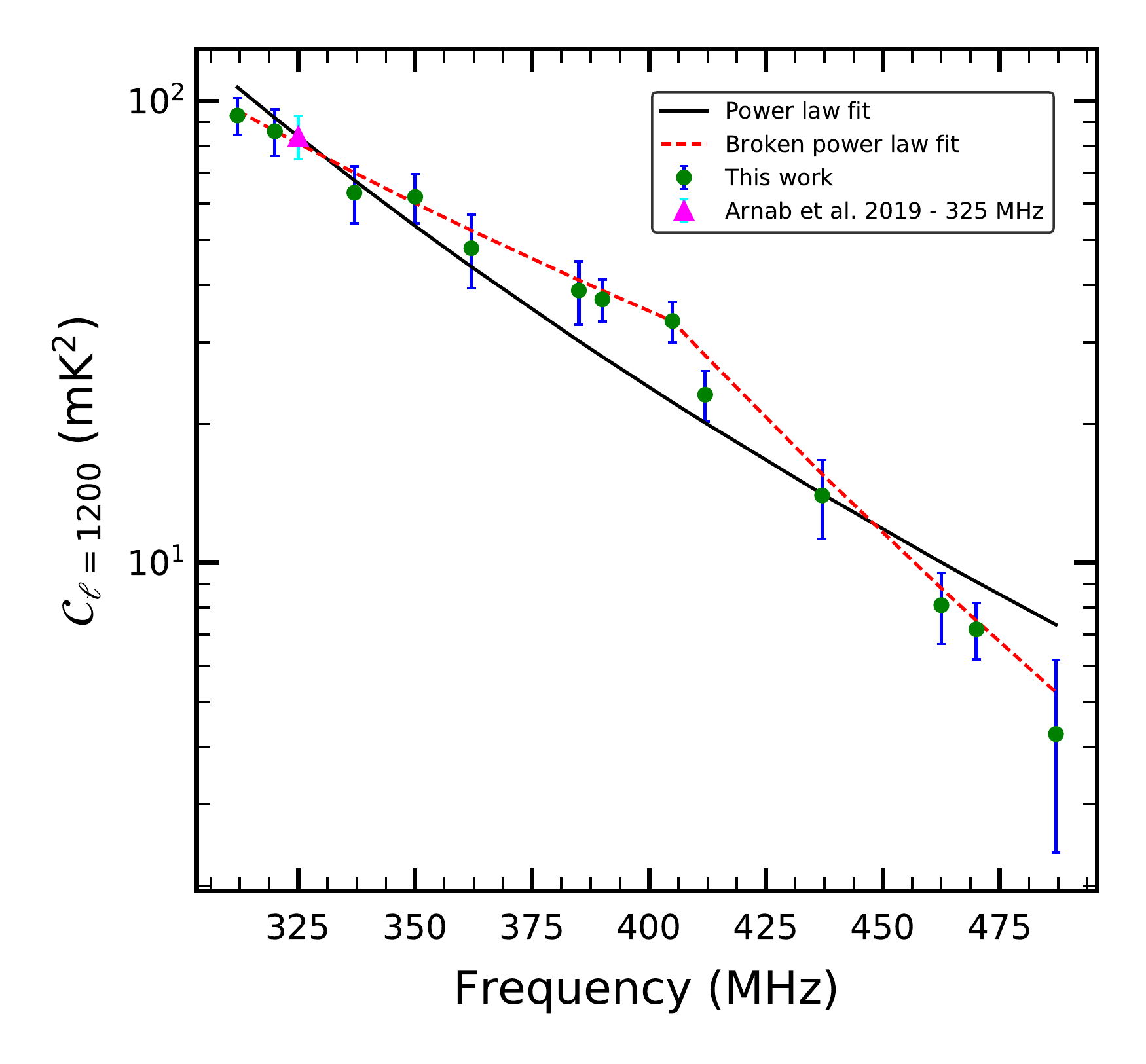}
    \caption{Angular power spectrum of DGSE normalized at $\textit{l}=1200$ as a function of frequency. The magenta triangle is the measured power spectrum of DGSE at 325 MHz (\citealt{Chakraborty2019}). The observed values are consistent with the previous measurement. }
    \label{cl_vs_f}
\end{figure*}

We have divided the  residual visibility data of whole bandwidth (200 MHz)  into 25 chunks of 8 MHz band. For each chunk of 8 MHz band, we have run TGE to estimate the angular power spectrum (2D).  This gives us  $C_{\mathcal{\ell}}^{i}$ (APS)  of  DGSE at the central frequency  of the $i^{th}$ chunk. Due to flagging and sparse $\textit{uv}$-coverage, we are able to estimate APS of DGSE  for only 13 chunks of residual visibilities. Then for each chunk we have found a $\mathcal{\ell}$ range where $C_{\mathcal{\ell}}^{i}$ shows a steep power law behavior, which is characteristics of DGSE (see Fig. \ref{power_plots}). We inferred that DGSE is dominant for that $\mathcal{\ell}$ range, beyond which residual point sources begins to dominate over DGSE. For that particular angular multipole ($\mathcal{\ell}$) range we fit a power law of the form : 
\begin{equation}
   \mathcal{C}_{\mathcal{\ell}}^{i} = A_{i} \mathcal{\ell} ^{ -\beta_{i}},
\end{equation}
where $A_{i}$ and $\beta_{i}$ are the amplitude and power law index of APS for the ith chunk. We have normalized the APS of all 13 chunks at $\mathcal{\ell}$ = $\mathcal{\ell}_{0}$ = 1200, i.e, $\mathcal{C}_{\mathcal{\ell} = 1200}^{i}$ = $A_{i}$. We have checked with other values of $\mathcal{\ell}_{0}$, but our findings are consistent. The value of $\beta$ for all 13 sub-bands lies between [1.8 to 3].  All  the plots of $\mathcal{C}_{\mathcal{\ell}}^{i}$ as a function of $\mathcal{\ell}$ for all 13 sub-bands are presented in the Fig. \ref{power_plots}.

 The values  of $\mathcal{C}_{\mathcal{\ell} = 1200}$ at the central frequency of 13 sub-bands ($\nu_{0}$) is being plotted in Fig. \ref{cl_vs_f}.  We have also plotted the measured value of the amplitude of  DGSE power spectrum at 325 MHz \citep{Chakraborty2019} in magenta. The spectral variation of APS for DGSE or the Multi-Frequency Angular Power Spectrum (MFAPS) of the DGSE is modelled as $\mathcal{C}_{\mathcal{\ell}}(\nu) \propto \nu^{-2\alpha}$.  Here, we have also fitted a  power law in frequency  to the whole frequency range given as : 

\begin{equation}
    \mathcal{C}_{\mathcal{\ell} = 1200} (\nu) =  A \nu ^{ -2\alpha}
\end{equation}
The value of $\alpha$ for whole frequency range is $2.9 \pm 0.21$. The reduced $\chi^{2}$ ($\chi^{2}_{R}$) value for this fit is  1.6. We have shown the fitted curve (black) in Fig. \ref{cl_vs_f}. 

Previously, \citet{Oliveira2008} presented a global sky model (GSM) of diffuse radio background using different total power large-area radio surveys between 10 MHz and 94 GHz. In their model (GSM), spectral index of diffuse emission  at 150 MHz is $\sim 2.6$ above the Galactic plane.  EDGES team has measured the spectral index of diffuse radio emission using all-sky averaged data. They have reported a mean  spectral index at high Galactic latitudes to be $2.52 \pm 0.04$ between frequency range  150 - 408 MHz \citep{Rogers2008}.  \citet{Mozdzen2017}   have found a spectral index nearly $2.62 \pm  0.02$ in frequency range 90-190 MHz using EDGES high-band system. Recently,  using EDGES low-band system, \citet{Mozdzen2019}  measured a spectral index lies between [2.54-2.59]  in frequency range 50-100 MHz. We have estimated the MFAPS of DGSE  power spectrum  for the first time with a wide-band radio interferometric observation. Our findings for ELAIS N1 field is consistent with previous total power observations. 

Since the reduced $\chi^{2}$ value for the single spectral index fit is high, we explored the possibility of a broken power law fit to the data as well with a break at 405 MHz ($\nu_{\mathrm{break}}$). The broken power law model is given by: 

\begin{equation}
    \mathcal{C}_{\mathcal{\ell} = 1200} (\nu) = \begin{cases} A \Big(\frac{\nu}{\nu_{\mathrm{break}}}\Big) ^{ -2\alpha_{1}}, \text{for $\nu < \nu_{\mathrm{break}}$}\\ 
     A \Big(\frac{\nu}{\nu_{\mathrm{break}}}\Big) ^{ -2\alpha_{2}}, \text{for $\nu > \nu_{\mathrm{break}}$}  
\end{cases}
\end{equation}
The best fitted values of spectral index for this case is $\alpha_{1} = 2.1 \pm 0.2$  and $\alpha_{2} = 4.8 \pm 0.4$.  The value of reduced $\chi^{2}$ is 0.3 for this broken power law fitting.

From the above two attempts to fit the MFAPS data with a broken or single power law, none of the models can be ruled out. The error bars in the MFAPS data makes it difficult to distinguish between both the models. Hence, a single spectral index of the MFAPS of DGSE cannot be ruled out. This is consistent with the findings so far with other radio telescopes and other parts of the sky. 

For the broken power law model, a break in the power law around 405 MHz suggests that there is a suppression of power above $\nu_{\mathrm{break}}$= 405 MHz and is due to ``synchrotron age''. The observed value of spectral index above the break ($\alpha_{2}$) is in between the JP (Jaffe-Perola) and   the KP (Kardashev-Pacholczyk) model \citep{Myers1985, Carilli1991}. The corresponding ``synchrotron age'' of the plasma is 80 Myr (using Eqn.1 of \citealt{Carilli1991}), assuming average magnetic field $B = 10 \mu$G.  A deeper analysis of spectral variation of the MFAPS requires more sensitive and much wider bandwidth data which is outside the scope of this paper.

\section{Conclusion}
\label{conclusion}

In this paper, we have shown deep observation of the ELAIS N1 field with the uGMRT at 300 - 500 MHz spanning a sky coverage of  $\sim$ 1.8 deg$^{2}$.  The field lies at high galactic latitude ($b=+44.48^{\circ}$) due to which it  helps us to study  extragalactic sources. Here we present the image of ELAIS N1 field and the catalog extracted from that image. The final image reaches an rms depth of $\sim$ 15 $\mu$Jy beam$^{-1}$ near the phase centre. The catalog presented here contains 2528 sources. 

We have discussed in detail the comparison of our catalog with previous radio catalogs  at other frequencies. We have found that flux scale is nearly consistent with other observations and the estimated ratio of flux densities of selected sample of sources when compared with other catalogs  are close to 1. 
We have also checked for astrometry after comparing with high frequency catalogs. The positional offset typically constrained within $\sim 0.5\arcsec$. This ensures the good agreement of positional information of radio sources with other radio catalogs. A well constrained positional accuracy is needed for identification of sources in optical catalogs which in turn helps us to study spectroscopic property of those sources. We have not shown cross matching with multi frequency data (other than radio) available for this field here. This defers to later work. Finally, we have estimated spectral indices after comparing flux densities with other low and high frequency radio catalogs covering the ELAIS N1 field. We have found a median spectral index of $\sim -0.7$ after comparing with 1.4 GHz NVSS and FIRST catalog and with low-frequency GMRT observations of the ELAIS N1 field (610 MHz GMRT).  A detailed investigation of spectral index using other frequency band data is deferred to late work. 

We also present the Euclidian-normalized source counts at 400 MHz after accounting for different correction factors. The corrected source counts are in good agreement with previous measurements at high flux densities. Similar to previous findings, we have also found a flattening in source counts  below 1 mJy. This flattening corresponds to increase in population of SFGs and radio-quiet AGNs. 

Finally, we have quantified the fluctuations in DGSE in this field and found out its evolution as a function of frequency. In general, DGSE is modelled as a simple power law both in angular and frequency domain. Although there is a hint of a broken power law in the MFAPS of DGSE, the sensitivity of the current observation prevents us from ruling out the single power law fit. Hence, more sensitive observations using much wider bandwidth data is required to infer conclusively. 

It should be noted that foreground modeling is critical for redshifted 21 cm signal experiments. Any errors in modeling the foregrounds can affect the detection of redshifted HI 21 cm signal. This study of spectral variation of the DGSE will facilitate to create more sensitive spectral and spatial models of the foreground, in particular the DGSE. This study also helps us to understand the foreground properties in this field and will be helpful for next generation telescopes such as the LOFAR, PAPER, HERA, SKA, which are trying to detect the  21 cm signal from  the EoR and post-EoR epoch. \\

{\bf ACKNOWLEDGEMENTS}

We  thank the staff of GMRT for making this observation possible. GMRT is run by National Centre for Radio Astrophysics of the Tata Institute of Fundamental Research. AC would like to thank DST for INSPIRE fellowship. AC thanks Ramij Raja, Anirban Roy and Manoneeta Chakraborty for helpful discussion. AC would like to thank Majidul Rahaman for making Tapered Gridded Estimator (TGE) parallel. NR acknowledges support from the Infosys Foundation through the Infosys Young Investigator grant.  We warmly thank the anonymous referee and the scientific editor for helpful comments and suggestions that have helped to improve this work.









\bsp	
\label{lastpage}
\end{document}